\documentclass{article}
\usepackage{svg}

\usepackage{microtype}
\usepackage{graphicx}
\usepackage{subcaption}
\usepackage{booktabs}

\usepackage{hyperref}

\usepackage[preprint]{icml2026}

\usepackage{amsmath}
\usepackage{amssymb}
\usepackage{mathtools}
\usepackage{amsthm}

\usepackage[capitalize,noabbrev]{cleveref}

\theoremstyle{plain}

\theoremstyle{definition}

\theoremstyle{remark}

\usepackage[textsize=tiny]{todonotes}

\icmltitlerunning{GRAB: An LLM-Inspired Sequence-First Click-Through Rate Prediction Modeling Paradigm}

\begin{document}

\twocolumn[
  \icmltitle{GRAB: An LLM-Inspired Sequence-First Click-Through Rate Prediction Modeling Paradigm}

  \icmlsetsymbol{equal}{*}

  \begin{icmlauthorlist}
    \icmlauthor{Shaopeng Chen}{baidu}
    \icmlauthor{Chuyue Xie}{baidu}
    \icmlauthor{Huimin Ren}{baidu}
    \icmlauthor{Shaozong Zhang}{baidu}
    \icmlauthor{Han Zhang}{baidu}
    \icmlauthor{Ruobing Cheng}{baidu}
    \icmlauthor{Zhiqiang Cao}{baidu}
    \icmlauthor{Zehao Ju}{baidu}
    \icmlauthor{Yu Gao}{baidu}
    \icmlauthor{Jie Ding}{baidu}
    \icmlauthor{Xiaodong Chen}{baidu}
    \icmlauthor{Xuewu Jiao}{baidu}
    \icmlauthor{Shuanglong Li}{baidu}
    \icmlauthor{Lin Liu}{baidu}
  \end{icmlauthorlist}

  \icmlaffiliation{baidu}{Baidu Inc., Beijing, China}

  \icmlcorrespondingauthor{Shuanglong Li}{lishuanglong@baidu.com}

  \icmlkeywords{Machine Learning, ICML}

  \vskip 0.3in
]

\printAffiliationsAndNotice{}  

\begin{abstract}
  Traditional Deep Learning Recommendation Models (DLRMs) face increasing bottlenecks in performance and efficiency, often struggling with generalization and long-sequence modeling. Inspired by the scaling success of Large Language Models (LLMs), we propose Generative Ranking for Ads at Baidu (GRAB), an end-to-end generative framework for Click-Through Rate (CTR) prediction. GRAB integrates a novel Causal Action-aware Multi-channel Attention (CamA) mechanism to effectively capture temporal dynamics and specific action signals within user behavior sequences. Full-scale online deployment demonstrates that GRAB significantly outperforms established DLRMs, delivering a 3.05\% increase in revenue and a 3.49\% rise in CTR. Furthermore, the model demonstrates desirable scaling behavior: its expressive power shows a monotonic and approximately linear improvement as longer interaction sequences are utilized.
\end{abstract}

\section{Introduction}
For a long time, Deep Learning Recommendation Models (DLRMs) \cite{naumov2019deep} have remained the mainstream choice in industrial recommender systems, especially for advertising Click-Through Rate (CTR) prediction \cite{mudigere2022software,zhou2018deep,cheng2016wide,guo2017deepfm,bai2025comprehensive,wang2017deep,ma2018modeling}, due to their strong capability to process high-cardinality sparse features and to model feature interactions with expressive neural networks. However, as user behavior data grows exponentially, traditional DLRMs face increasing bottlenecks in both \emph{performance} and \emph{efficiency} (detailed discussions in Appendix~\ref{sec:performance_efficency}). Fundamentally, DLRMs rely on rule-based feature engineering and suffer from the inherent flaw of ``strong memory, weak reasoning'' \cite{cheng2016wide,wu2024survey}. They often fail to generalize to new ads or scenarios that require logical inference, and their gains exhibit diminishing returns: further improvements typically demand exponentially increasing computational costs, rendering long-term deployment and iteration economically unsustainable \cite{zhang2024scaling,mudigere2022software}.

Departing from the structural constraints of DLRMs, the rise of Large Language Models (LLMs) has been driven by scaling laws \cite{kaplan2020scaling,zhang2024wukong,zhang2024scaling}, where performance predictably improves with increased parameters, data, and compute. This success has inspired the extension of scaling laws to recommendation systems, fostering the LLMs for Recommendation (LLM4Rec) paradigm \cite{wu2024survey,li2024large}(see Appendix~\ref{sec:LLM4rec} for a taxonomy). A key innovation within this framework is Generative Recommendation (GR)\cite{li2024large,rajput2023recommender}. Representative works like HSTU \cite{zhai2024actions} formulate recommendation as autoregressive sequence prediction, effectively modeling long user sequences to capture temporal dynamics \cite{zhou2018deep,zhou2019deep,kang2018self,sun2019bert4rec}. Crucially, GR exhibits scaling properties similar to LLMs, offering a practical path to transcend the performance bottlenecks of traditional DLRMs \cite{naumov2019deep,zhai2024actions,zhang2024wukong}.

Despite these theoretical advancements, deploying GR models in high-throughput industrial systems remains challenging due to strict online serving and optimization constraints. The primary obstacle is \emph{computational efficiency}. Standard Transformer training requires extensive padding for variable-length sequences, resulting in significant computational waste \cite{vaswani2017attention,krell2021efficient}. While sequence packing---a common Natural Language Processing (NLP) technique for concatenating multiple short sequences---effectively mitigates this issue \cite{krell2021efficient}, its straightforward application to recommendation systems triggers a more subtle yet damaging failure mode: \emph{Distribution Skew} \cite{baylor2017tfx,polyzotis2019data,sculley2015hidden,han2025mtgr}.

In recommendations, packing a user's full history creates mini-batches with excessive intra-user correlation, which violates the i.i.d. assumption typically relied on by SGD-style optimization \cite{doan2020finite}. This skew (details in Appendix~\ref{appendix:skew}) causes sparse parameters (i.e., embedding tables) to overfit specific users, hindering the generalization of dense parameters (e.g., Transformer weights responsible for inference) \cite{naumov2019deep,li2024embedding}. This reveals a fundamental tension: sparse parameters require diverse, uncorrelated samples for robust ``memorization'', whereas dense parameters benefit from long, coherent contexts for sequential ``reasoning'' \cite{cheng2016wide,kang2018self,sun2019bert4rec}. This misalignment implies that standard synchronous training on packed sequences may lead to suboptimal convergence due to the conflicting gradient requirements of the sparse and dense components \cite{yu2020gradient}.

Meanwhile, existing GR models typically ignore data heterogeneity, resulting in performance limitations (see Appendix~\ref{appendix:lack_of_performance} for detailed discussion). To overcome these challenges, we propose Generative Ranking for Ads at Baidu (GRAB), an end-to-end sequential training and inference framework for industrial-grade CTR prediction. GRAB introduces three core innovations to reconcile the demands for performance, efficiency, and training stability:

\begin{itemize}
\item \textbf{End-to-End Framework}: We introduce GRAB, an end-to-end framework that combines the strengths of DLRMs and GR. Specifically, it fuses the large-scale sparse feature engineering inherent to DLRMs with the sequential inference capabilities of GR, thereby achieving a balance between explicit memorization and implicit reasoning.
\item \textbf{Causal Action-aware Multi-channel Attention (CamA) mechanism}: We propose CamA, a multi-channel, action-aware mechanism designed to boost model performance by modeling both user exposure and interaction signals, improving generalization and robustness across tasks and scenarios.
\item \textbf{Sequence-Then-Sparse (STS) Training}: To address distribution skew from sequence packing, we propose STS, a training strategy that decouples the optimization of dense parameters and sparse embeddings. This resolves their gradient conflict, stabilizes training, and improves convergence without extra compute, enabling high-throughput industrial deployment of GR.
\end{itemize}

We have completed a comprehensive evaluation of GRAB in Baidu's commercial advertising CTR ranking business. In offline comparisons, GRAB outperformed  mainstream industrial DLRMs as well as emerging GR models (achieving 0.19\% relative improvement over the best baseline). Compared to the production DLRM baseline, GRAB achieved an AUC uplift of approximately 2 basis points in online A/B testing, resulting in a 3.05\% increase in CPM and a 3.49\% increase in CTR. Furthermore, scaling analysis demonstrates that the model’s AUC improves monotonically with both model capacity and the length of behavior sequences, indicating that the architecture can stably benefit from modeling longer behavior chains without saturation.

\section{Related Works}
\label{sec:related}
\textbf{DLRM-based industrial CTR prediction.} Industrial CTR prediction has long been dominated by DLRMs, which embed high-cardinality categorical fields and model feature interactions via MLP/Cross-style modules \cite{naumov2019deep,cheng2016wide,guo2017deepfm,wang2017deep}. To incorporate user histories, production systems often attach explicit behavior encoders to DLRMs, e.g., target-attention/memory based models such as DIN, DIEN, MIMN, and SIM \cite{zhou2018deep, zhou2019deep, pi2019practice, pi2020search}, as well as stronger industrial variants like TWIN \cite{si2024twin}. Despite their effectiveness, these approaches still heavily rely on hand-crafted statistics and engineered cross features \cite{he2014practical,cheng2016wide}, and typically compress long histories into fixed-size vectors, making it difficult to scale to long sequences and heterogeneous action signals \cite{pi2019practice}.

\textbf{GR.} Recent GR work models recommendation as causal Transformer-based sequential prediction, enabling long-context modeling and exhibiting favorable scaling behavior \cite{zhai2024actions,chai2025longer,petrov2023generative}. However, deploying GR in the industrial advertising CTR stack still presents challenges in the following aspects: (i) bridging large-scale sparse feature engineering with tokenized sequential modeling \cite{han2025mtgr, chai2025longer}, (ii) modeling heterogeneous action semantics often discarded by naive homogeneous serialization \cite{zhai2024actions}, and (iii) training instability introduced by sequence packing (distribution skew) under strict optimization constraints \cite{krell2021efficient}. Detailed discussion of GR and comparisons between GRAB and related work are given in Appendix~\ref{sec:appendix_gr}.

\section{Methodology}

\subsection{DLRMs}
\label{sec:dlrm}
\begin{figure}[htbp]
    \centering
    \includegraphics[width=0.70\linewidth]{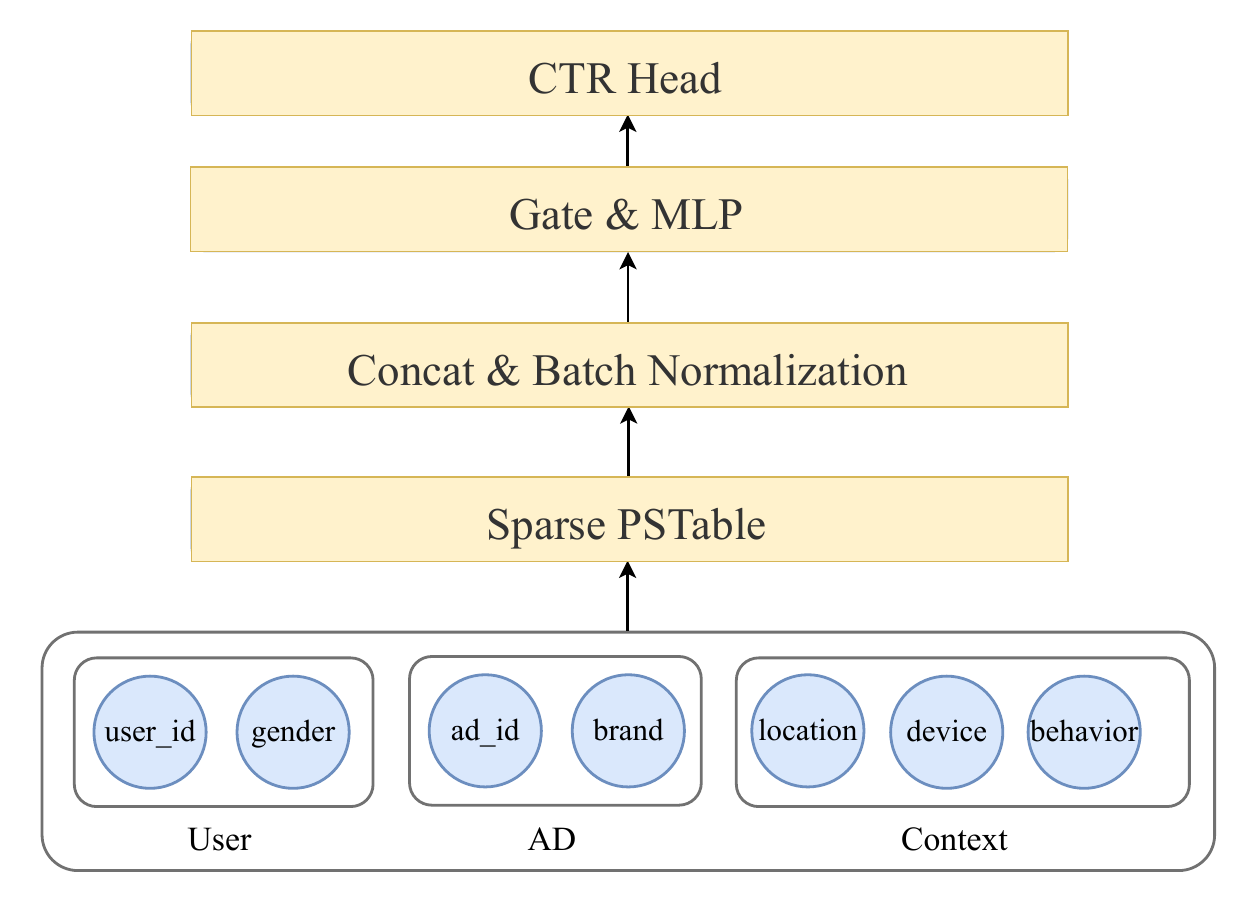}
    \caption{The traditional DLRM architecture: sparse features are hashed to IDs and embedded via PSTable, and then concatenated into a fixed-length flattened vector for CTR prediction.}
    \label{fig:dlrm}
\end{figure}

The traditional DLRM architecture, as shown in Fig.~\ref{fig:dlrm}, follows a modular processing pipeline for CTR prediction, handling raw features from users, candidate ads, and contextual signals. The pipeline involves: (a) expanding categorical features into fixed fields via feature engineering, (b) mapping these fields through hashing to obtain discrete ID vectors for embedding lookup in a Sparse Parameter Server Table(PSTable), and (c) concatenating and normalizing the retrieved embeddings to form a fixed-length flattened vector. This unified representation is then fed into an MLP, typically enhanced with a gating network, to model high-order feature interactions and generate the final CTR prediction.

\subsection{Overall Architecture of GRAB}
\label{sec:overall}
\begin{figure*}[t]
    \centering
    \includegraphics[width=0.8\linewidth]{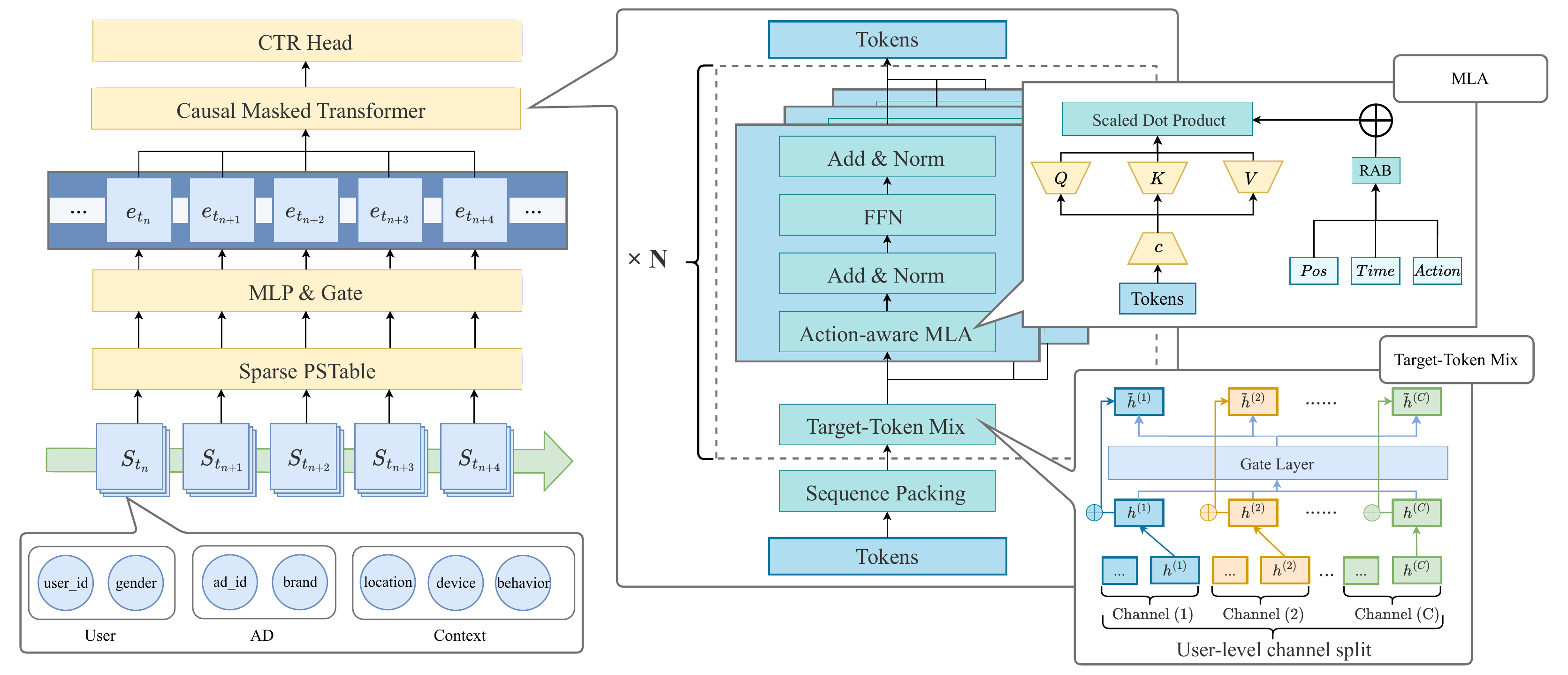}
    \caption{Overview of GRAB’s end-to-end CTR prediction pipeline: (1) Tokenizing raw fields via a sparse PSTable and fusing them into event tokens. (2) Packing tokens per user with causal and heterogeneous masks. (3) Processing through $N$ Transformer layers equipped with the Causal Action-aware Multi-channel Attention (CamA) mechanism. (4) Final CTR prediction from the output representations.}
    \label{fig:figure_grab}
\end{figure*}

GRAB, with the overall architecture shown in Fig.~\ref{fig:figure_grab}, is designed to model user behavior history sequences in an end-to-end manner, as applied in scenarios like CTR prediction. GRAB follows a three-stage pipeline: (i) \emph{sparse feature layer}; (ii) \emph{dense tokenizer}; and (iii) \emph{sequence modeling layer}. 
Given raw behavior logs, GRAB first converts heterogeneous categorical signals into sparse IDs at the event level, then tokenizes each event into a dense representation, and finally applies a sequence model to estimate the click probability of candidate ads. GRAB uses its dense representation calculated from the dense tokenizer to bridge DLRM-style sparse feature engineering and GR-style sequential modeling, enabling end-to-end training and inference along a single, unified computation path from input to output, thereby improving CTR prediction performance through end-to-end sequential modeling of event-level user behaviors. 

\textbf{Sparse Feature Layer.} The sparse feature layer (details in Appendix~\ref{sec:sparse}) processes raw logs into time-ordered event sequences. Each event's categorical fields are converted into sparse IDs using standard DLRM feature engineering (Section~\ref{sec:dlrm}), yielding a structured sequence of events annotated with field-wise IDs.

\textbf{Dense Tokenizer.} 
Unlike DLRM, which collapses field embeddings into a fixed-length, order-agnostic vector for pointwise processing, GRAB preserves the temporal event structure. It aggregates per-event field embeddings and projects them into $\mathbb{R}^{d_\text{model}}$ to form sequential event tokens (Appendix~\ref{sec:dense}), resulting in a time-ordered token sequence. This sequence serves as the input to a subsequent Transformer, thereby enabling the modeling of long-range dependencies and interest drift.

\paragraph{Autoregressive-like Sequence Modeling Layer.}

Built on sequence packing (Section~\ref{causal_mask}), heterogeneous tokens (Section~\ref{sec:het}),  and action-aware relative attention bias (Section~\ref{sec:RAB}), our core contribution is the \textbf{CamA} mechanism (Section~\ref{sec:multi_channel}). CamA integrates a multi-channel design for parallel processing of diverse behaviors and inherits action-aware contextualization from RAB, providing a unified and efficient framework for modeling complex user interest patterns across scenarios.

\subsection{Autoregressive-like Sequence Modeling Layer}
\label{sec:layer_sequence}
Following the dense tokenizer, this layer is designed to capture the temporal dependencies and dynamic evolution of user interests, which takes the sequence of dense event tokens generated by the preceding layer as input (as described in Appendix~\ref{sec:appendix_seq}). Formally, for a user $u$, the input sequence consists of the behavior history $\mathbf{E}^{\text{beh}} = \{\mathbf{e}^{\text{beh}}_t\}_{t=1}^{T_u}$ and the candidate advertisements $\mathbf{E}^{\text{ad}} = \{\mathbf{e}^{\text{ad}}_i\}_{i=1}^{N_u}$, where $\mathbf{e}^{\text{beh}}_t, \mathbf{e}^{\text{ad}}_i \in \mathbb{R}^{d_{\text{model}}}$ are the dense embeddings of the $t$-th behavior event and the $i$-th candidate ad, respectively, $T_u$ is the behavior history length, and $N_u$ is the number of candidate ads.

\begin{figure*}[t]
    \centering
    \begin{subfigure}[b]{0.6\linewidth}
        \centering
        \includegraphics[width=\linewidth]{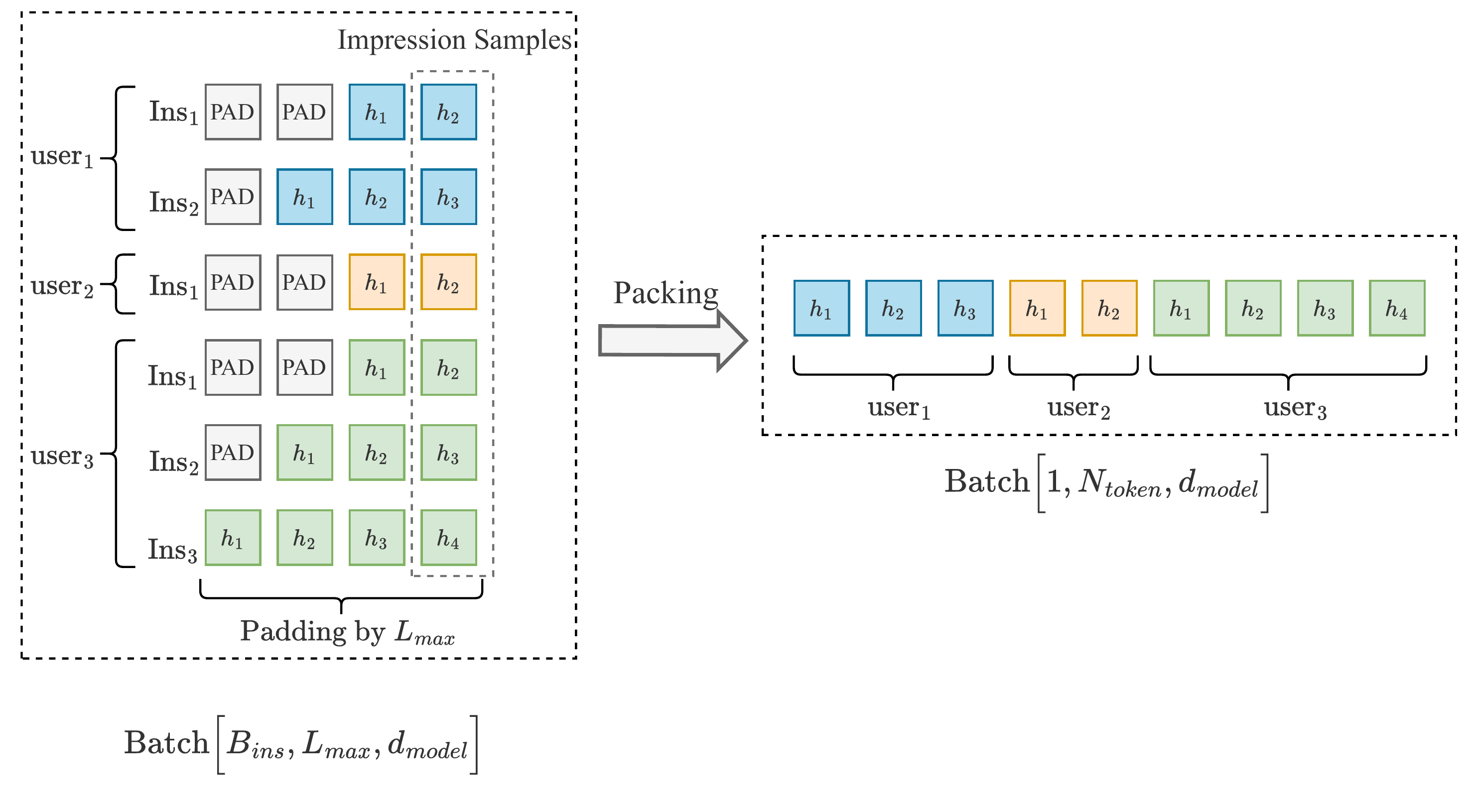}
        \caption{Sequence Packing}
        \label{fig:sequence_packing}
    \end{subfigure}
    \hfill
    \begin{subfigure}[b]{0.35\linewidth}
        \centering
        \includegraphics[width=\linewidth]{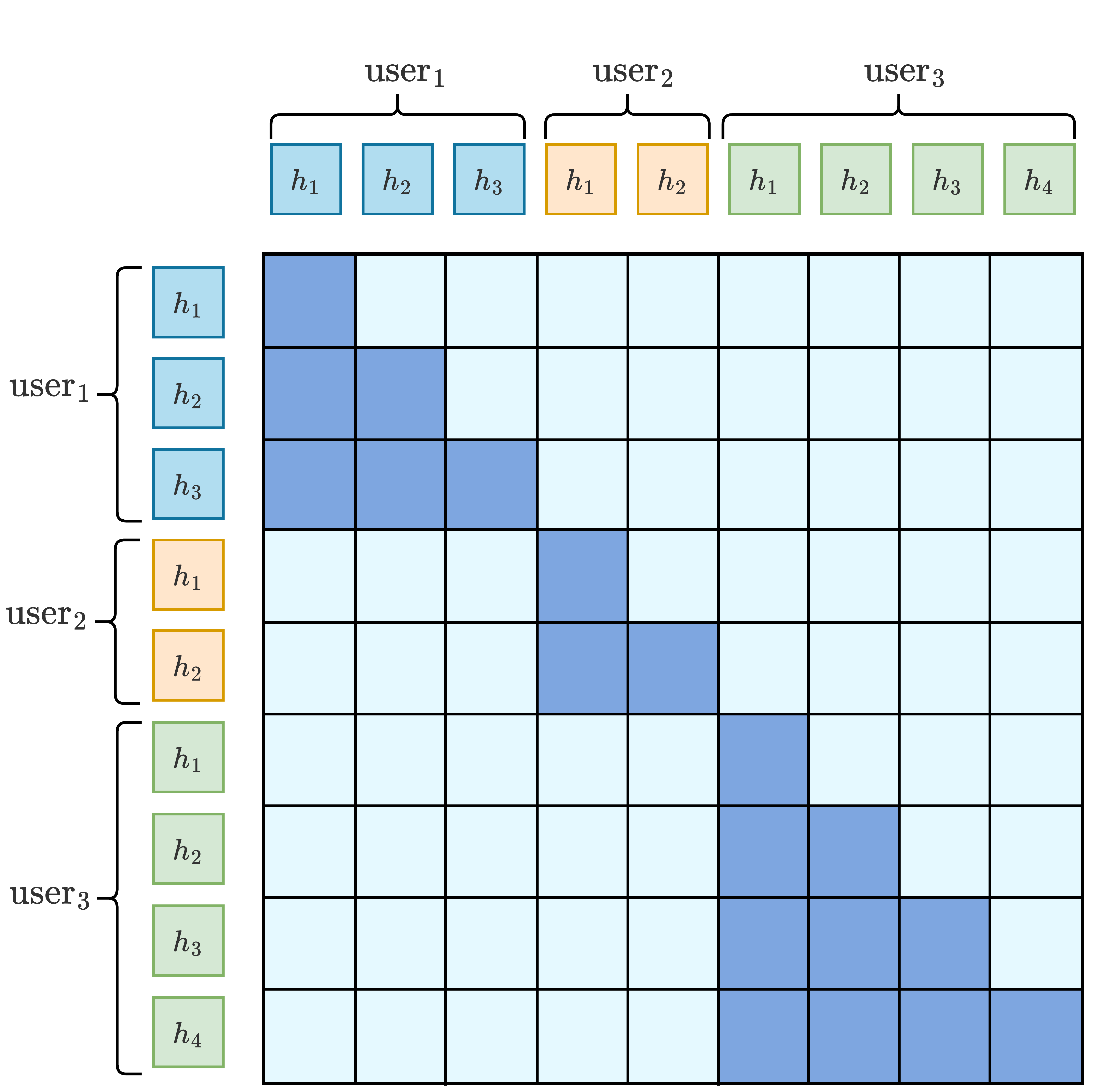}
        \caption{User-isolated Causal Mask}
        \label{fig:attention_mask}
    \end{subfigure}
    \caption{Sequence packing and user-isolated causal masking in GRAB. (a) Instead of padding each impression instance to a fixed length $L_{\max}$, tokens from multiple impressions are concatenated within each user and different users are kept in disjoint segments, yielding a single packed sequence of length $N_{token}$ for compute-efficient batching. (b) The user-isolated causal mask exhibits a block-diagonal lower-triangular pattern, so each token can only attend to past tokens within the same user segment, enforcing both user isolation and temporal causality.}
    \label{fig:combined_figures}
\end{figure*}

\subsubsection{Sequence Packing and User-isolated Causal Mask}
\label{causal_mask}

In industrial training logs, as shown in the left image of Fig.~\ref{fig:sequence_packing}., a mini-batch is typically formed by sampling $B_{\mathrm{ins}}$ impression instances.
Each instance contains a variable-length token sequence composed of (i) a subsequence of the user's historical behavior tokens and (ii) target advertisement tokens to be scored. A straightforward batching strategy pads every instance to a fixed length $L_{\max}$, yielding a dense tensor with dimensions $B_{\mathrm{ins}}$$\times$$L_{\max}$$\times$$d_{\mathrm{model}}$,

which introduces substantial computational waste when most instances are much shorter than $L_{\max}$.

To eliminate such padding overhead while preserving the temporal semantics, GRAB performs sequence packing by grouping tokens by user. Specifically, tokens from multiple impression instances belonging to the same user $u$ are merged into a single contiguous token segment, while segments of different users are strictly separated.
Within each user segment, all tokens are stably sorted by timestamp so that the packed segment forms a single timeline for sequential modeling. After packing, the batch is represented as one long packed tensor $H = \text{Pack}(\mathbf{E}^{beh}, \mathbf{E}^{ad}) \in \mathbb{R}^{1 \times L \times d_{\text{model}}}$, where $L$ denotes the total packed length across all users in the mini-batch.

For convenience, we associate each packed position $p\in\{1,\dots,L\}$ with
(i) a \emph{segment id} $\sigma(p)\in U_B$ indicating which user it belongs to,
and (ii) a \emph{local time index} $\ell(p)\in\{1,\dots,L_{\sigma(p)}\}$ within that user segment.

\textbf{User-isolated causal mask.} On the packed tensor $H$, we construct an additive attention mask $M^{\mathrm{pack}}\in\mathbb{R}^{L\times L}$ that enforces two constraints:
(1) \emph{user isolation} (no cross-user attention), and
(2) \emph{causality} within each user's timeline (no future leakage).
Formally, for query position $p$ and key position $q$,
\begin{equation}
M^{\mathrm{pack}}_{p,q}=
\begin{cases}
1, & \text{if } \sigma(p)=\sigma(q)\ \text{and}\ \ell(q)\le \ell(p),\\
0, & \text{otherwise}.
\end{cases}
\label{eq:mpack}
\end{equation}
This yields a block-diagonal lower-triangular structure(as shown in Fig.~\ref{fig:attention_mask}), where each block corresponds to one user segment.

\subsubsection{Heterogeneous Behavior Tokens and Heterogeneous Visibility Mask}
\label{sec:het}
After sequence packing, for each user $u$, we obtain a user-isolated, time-ordered packed stream with its causal mask $M^{\mathrm{pack}}$. To further reduce redundancy in the packed history while preserving the information needed for scoring the current candidate, we instantiate two token views at each packed timestamp $t$: \textbf{the partial token (history)} $h_t \in \mathbb{R}^{d_{\mathrm{model}}}$, which retains only time-varying information that is useful for representing history and discards static or highly repetitive fields (e.g., user\_id) that would otherwise be duplicated across historical steps and could lead to overfitting; and \textbf{the full token (candidate)}  $h'_t \in \mathbb{R}^{d_{\mathrm{model}}}$, which retains the complete information required to score the candidate at time $t$, including the static fields omitted from the partial history view. We then interleave them to form the heterogeneous packed sequence:
$
H_u=\big[\mathbf{h}_1,\mathbf{h}'_1,\mathbf{h}_2,\mathbf{h}'_2,\ldots,\mathbf{h}_{T_u},\mathbf{h}'_{T_u}\big].
$

\begin{figure}[htbp]
    \centering
    \includegraphics[width=0.42\linewidth]{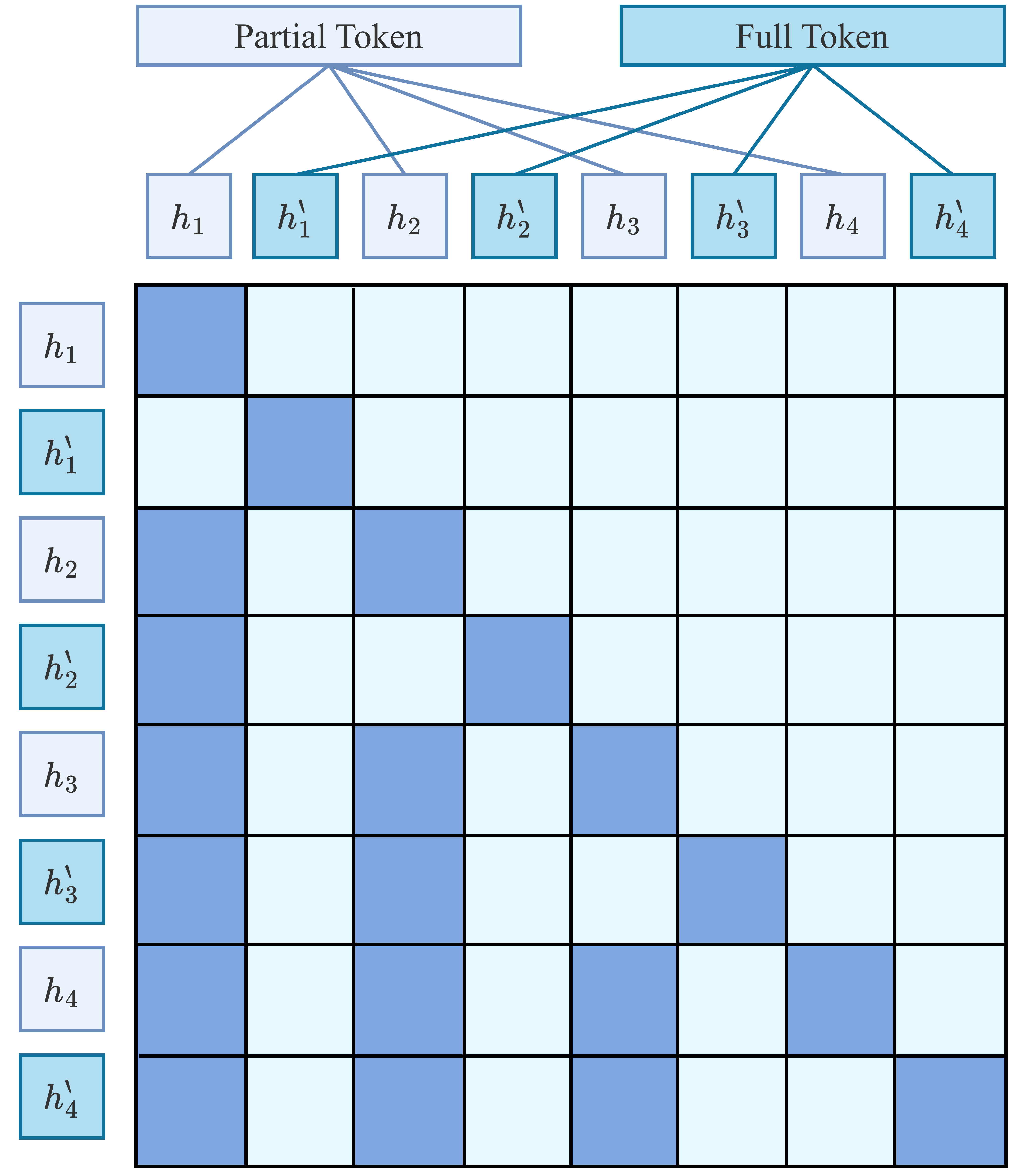}
    \caption{Heterogeneous behavior tokens and heterogeneous visibility mask $M^{\mathrm{het}}$ (blue entries). Partial tokens attend only to partial-history tokens up to the current time, while full tokens attend to partial-history tokens up to their time index and to themselves, but never to other full tokens, preventing duplicated static information from propagating along time.}
    \label{fig:het_mask}
\end{figure}

\textbf{Heterogeneous Visibility Mask.} On top of the user-isolated causal constraint encoded by $M^{pack}$, we apply a mask-rewriting operator $\mathcal{R}(\cdot)$ to obtain the heterogeneous visibility mask $M^{het}$. Concretely, $\mathcal{R}(\cdot)$ rewrites the visibility pattern according to the token types in the following way:
(i) partial ($\mathcal{P}$) tokens only attend to partial history tokens; and
(ii) full ($\mathcal{F}$) tokens attend to partial history tokens and themselves, but never attend to other full tokens. Formally, index positions in $H_u$ by $n\in\{1,\dots,2T_u\}$, we define the time index $\tau(n) = \lceil n/2 \rceil$ and token type $\kappa(n) = \mathcal{P}$ if $n$ is odd, otherwise $\kappa(n) = \mathcal{F}$.
Then the heterogeneous mask (as shown in Fig.~\ref{fig:het_mask}) is

\begin{equation}
M^{\mathrm{het}}_{p,q}=
\begin{cases}
1, & \kappa(p)=\mathcal{P},\ \kappa(q)=\mathcal{P},\ \tau(q)\le \tau(p),\\[2pt]
1, & \kappa(p)=\mathcal{F},\ \kappa(q)=\mathcal{P},\ \tau(q)\le \tau(p),\\[2pt]
1, & \kappa(p)=\mathcal{F},\ p=q,\\[2pt]
0, & \text{otherwise}.
\end{cases}
\label{eq:het_mask}
\end{equation}

\subsubsection{Action-aware Attention: Relative Encoding and Efficient Computation}
\label{sec:RAB}
On top of the heterogeneous behavior tokens and the heterogeneous visibility mask $M^{het}$, we further adopt a action-aware RAB(i.e., relative attention bias) causal attention mechanism. It augments standard multi-head self-attention with three designs: a causal mask to prevent future leakage, a dual sliding-window visibility constraint to support streaming-style training, and a query-aware relative bias that enables the query to directly interact with relative position/time/action signals.

\paragraph{Action-aware relative attention logits.}
Given a query $q_i$ and a key $k_j$, the attention logit is computed as
\begin{equation}
w_{i,j} = q_i^{\top}\cdot \left(k_j + Pos_{i,j} + Action_{i,j} + Time_{i,j}\right),
\label{eq:rab_logit}
\end{equation}
where $Pos_{i,j}$, $Action_{i,j}$, and $Time_{i,j}$ are learnable embeddings derived from relative position, relative action, and relative time, respectively. For continuous or large-range signals (e.g., action statistics or play durations), we first discretize them into buckets and then perform embedding lookup.

Compared with a query-agnostic relative bias (e.g., $w_{i,j} = q_i^{\top}k_j + Pos_{i,j} + \cdots$), Eq.~\ref{eq:rab_logit} makes the relative signals action-aware via the interaction $q_i^{\top}Pos_{i,j}$, $q_i^{\top}Action_{i,j}$, and $q_i^{\top}Time_{i,j}$, allowing the model to adaptively emphasize different contextual relations under different queries (i.e., target ads).

\paragraph{Causal mask with dual sliding windows.}

We enforce causality and further restrict attention using combined time and length windows. The mask is defined as $M^{\mathrm{rab}}_{p,q}=1$ if $q \le p$ and the distance $p-q$ does not exceed the length sliding-window limit $L_{\mathrm{w}}$; otherwise $M^{\mathrm{rab}}_{p,q}=0$.

This serves two key industrial purposes: (1) it bounds per‑token computation, guaranteeing stable throughput/latency over growing behavior histories; (2) it matches the online training paradigm—events arrive incrementally, and the model updates attention context on the fly without reprocessing the full sequence, boosting training efficiency and serving practicality.

\paragraph{Efficient computation.}
The naive implementation of Eq.~\ref{eq:rab_logit} would yield an $O(L^2 d_{\text{model}})$ intermediate tensor, which is prohibitively memory-intensive in practice. We adopt the optimization in~\cite{golovneva2024contextual} to reorder the computation. We define codebooks $B^{\text{pos/act/time}} \in \mathbb{R}^{N_\ast \times d_{\text{model}}}$ and bucketized indices $p_{i,j}, a_{i,j}, t_{i,j}$. Then Eq.~\ref{eq:rab_logit} can be equivalently written as:
\begin{equation}
w_{i,j}
= q_i^{\top}k_j
+ (s_i^{pos})\!\left[p_{i,j}\right]
+ (s_i^{act})\!\left[a_{i,j}\right]
+ (s_i^{time})\!\left[t_{i,j}\right].
\label{eq:rab_logit_opt}
\end{equation}
where $s_i^{\mathrm{pos}} = q_i^{\top}B^{\mathrm{pos}}$, $s_i^{\mathrm{act}} = q_i^{\top}B^{\mathrm{act}}$, and $s_i^{\mathrm{time}} = q_i^{\top}B^{\mathrm{time}}$. 
In practice, we first compute the projection vectors $s_i^{\ast}$, then obtain relative terms via fast gather operations. This completely avoids the large $L \times L \times d_{\text{model}}$ tensor, dramatically reducing peak memory and improving computational efficiency.

\begin{figure}[htbp]
    \centering
    \includegraphics[width=0.95\linewidth]{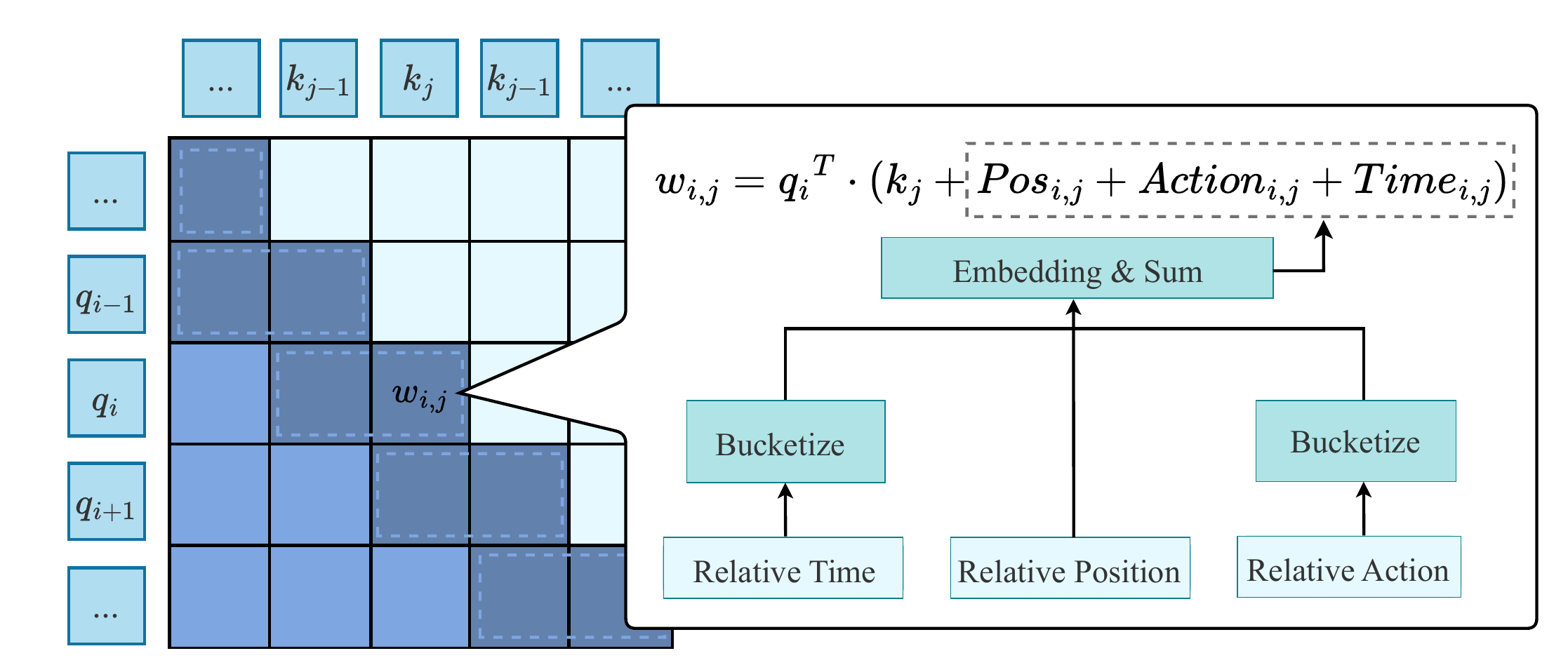}
    \caption{Action-aware relative attention bias (RAB) with efficient computation. Left: a causal mask with dual sliding windows, which limits each query to attend only to recent past tokens visible within the sliding-window. Right: the action-aware relative encoding pipeline: relative time, position, and action signals are bucketized (as needed), embedded, summed, and injected to the attention logits.}
    \label{fig:RAB}
\end{figure}

\subsubsection{Multi-channel Attention}
\label{sec:multi_channel}
While the action-aware RAB attention (Section~\ref{sec:RAB}) enhances each individual attention logit with relative position/action/time signals, it still treats the packed stream as a single mixed sequence. However, in industrial logs, user behaviors are highly heterogeneous (e.g., spanning different time windows or encompassing different behavior types), and different behavioral subsets often exhibit distinct temporal dynamics and predictive value. A straightforward design is to flatten all tokens into a single sequence and apply causal self-attention, yet this couples heterogeneous sources into one interaction graph and incurs a quadratic cost (e.g., $O((n+m)^2)$ for two sources with lengths $n$ and $m$).  To improve both modeling effectiveness and efficiency, we further introduce the \textbf{Causal Action-aware Multi-channel Attention (CamA)} mechanism, which integrates a multi-channel design, conceptually analogous to multi-head attention but with channel-specific visibility constraints. We therefore model each channel with an independent causal self-attention stack, and fuse the channel-wise representations via a lightweight gated mixing module. Let $\mathcal{C}=\{1,\dots,C\}$ denote the channel set. For each user, channel $c$ provides a token sequence
$\mathbf{X}^{(c)}\in\mathbb{R}^{T_c\times d}$, and we append the shared target token $X^{ad}\in\mathbb{R}^{d}$:
\begin{equation}
\mathbf{S}^{(c)} = [\mathbf{X}^{(c)};\mathbf{x}^{\mathrm{tar}}]\in\mathbb{R}^{(T_c+1)\times d},
\qquad t^\star = T_c+1.
\end{equation}

Each channel is equipped with its own causal visibility mask $\mathbf{M}^{(c)}$, and is encoded independently:
\begin{equation}
\begin{gathered}
\mathbf{H}^{(c,\ell+1)}
= \mathrm{Layer}^{(c)}_\ell\!\left(\tilde{\mathbf{H}}^{(c,\ell)}; \mathbf{M}^{(c)}\right),\\
\qquad
\mathbf{H}^{(c,0)}=\mathbf{S}^{(c)},\quad c\in\mathcal{C}.
\end{gathered}
\label{eq:mc_intra}
\end{equation}

\paragraph{Target-token gated mixing.}
To enable cross-channel information sharing while keeping computation lightweight, we perform mixing only on the target position $t^\star$ at each layer. The mixed representation $\tilde{\mathbf{h}}^{(c,\ell)}$ is obtained by first computing channel‑wise gating weights $\boldsymbol{\beta}^{(c,\ell)}$ and then aggregating information from all other channels:

\begin{equation}
\tilde{\mathbf{h}}^{(c,\ell)}
= \mathbf{h}^{(c,\ell)} + \sum_{i \in \mathcal{C} \setminus \{c\}} \beta^{(i,\ell)} \odot \mathbf{h}^{(i,\ell)}.
\label{eq:mc_mix}
\end{equation}
This updated representation replaces $\mathbf{h}^{(c,\ell)}$ at position $t^\star$, forming the updated channel representation $\tilde{\mathbf{H}}^{(c,\ell)}$ used in \eqref{eq:mc_intra}.
Finally, the concatenated last-layer target representations from all channels are used for CTR prediction.

\subsection{Sequence Then Sparse Training}

While sequence packing (Section~\ref{causal_mask}) significantly enhances computational efficiency, it introduces a critical challenge: distribution skew. Since samples within a packed mini-batch belong to the same user, the high intra-user correlation leads to redundant updates for specific sparse IDs, causing the model to overfit to specific user-ad interactions, rather than learning generalizable patterns. To mitigate this, we propose the Sequence Then Sparse (STS) training paradigm (detailed discussions in Appendix~\ref{appendix:STS}), a two-stage decoupled optimization strategy that balances long-range sequential modeling with robust sparse feature learning.

\subsubsection{Stage I: Sequence Modeling (Sequence Phase)}
The first stage focuses on capturing the evolution of user interests and temporal dependencies. We perform end-to-end autoregressive-like learning on the packed user sequences $Z$, which include candidate tokens and their historical trajectories. In this phase, we optimize the dense tokenizer and the causal Transformer, while keeping the Sparse Embedding Table $\Phi$ frozen. By freezing $\Phi$, we stabilize the token space, forcing the Transformer to focus exclusively on the relational dynamics between events rather than over-memorizing specific ID features.

\subsubsection{Stage II: Sparse Feature Learning (Sparse Phase)}
The second stage is designed to refine the discrete representations, particularly for long-tail IDs. In this phase, we revert to a non-sequential format, treating each sample as an independent user-ad exposure to break the distribution skewness. This stage optimizes the sparse embeddings $\Phi$, which act as a robust corrector for the gradient accumulation amplified by sequence packing. It ensures that the basic feature representations remain accurate and unbiased across the entire traffic distribution.

\section{System Deployment}
GRAB has been successfully deployed in a large-scale feed ad ranking system, handling billions of daily requests. Unlike conventional memory-bound DLRMs, GR is markedly compute-bound due to the quadratic complexity of Transformer self-attention. To satisfy stringent latency requirements, we implemented a co-designed hardware-software architecture. Due to space constraints, we provide the comprehensive system overview (Fig.~\ref{fig:pipeline}) and detailed deployment optimizations in Appendix~\ref{sec:appendix_deployment}.

\section{Experiment}

\subsection{Overall Performance Comparison}
\begin{figure*}[htbp]
    \centering
    \begin{subfigure}[b]{0.40\textwidth}
        \centering
        \includegraphics[width=1.0\textwidth]{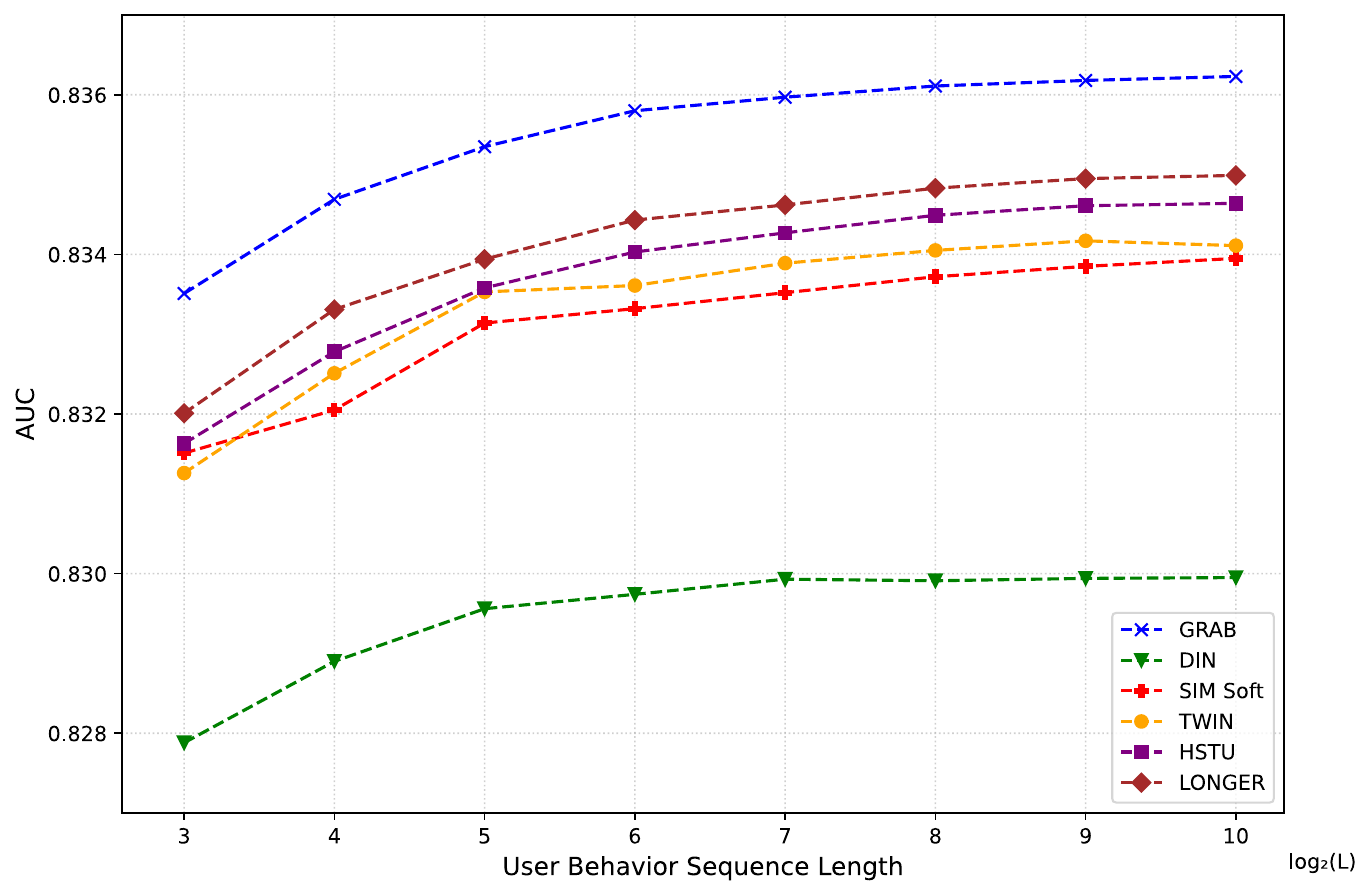}
        \caption{Overall Performance}
        \label{fig:overall}
    \end{subfigure}
    \hfill
    \begin{subfigure}[b]{0.40\textwidth}
        \centering
        \includegraphics[width=1.0\textwidth]{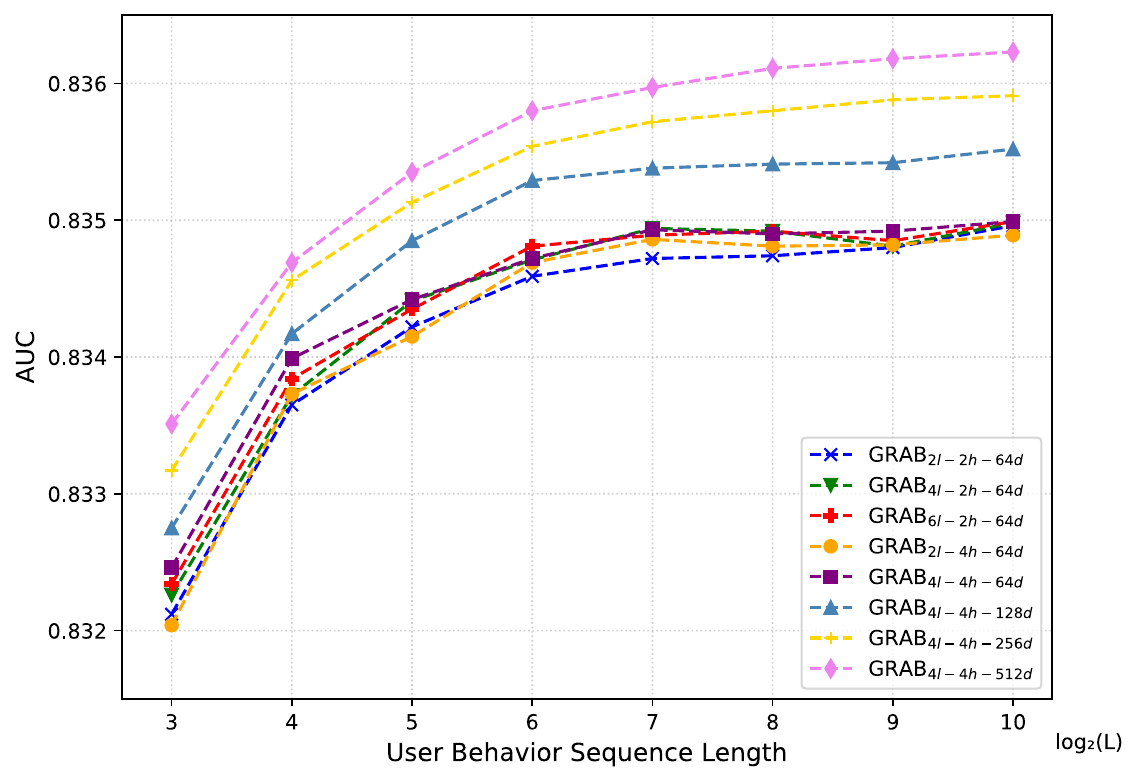}
        \caption{Scaling Performance}
        \label{fig:scaling}
    \end{subfigure}
    \caption{DLRMs vs. GRs across different user behavior sequence lengths (a), with a +0.1\% improvement in AUC, indicating a significant enhancement. GRABs comparison in different parameter scale(b)}
    \label{fig:exp_by_seq_len}
\end{figure*}

We first compared the performance of GRAB against state‑of‑the‑art recommendation models on the Baidu real‑world industrial dataset. The training data, derived from the Baidu real recommendation advertising scene, contains billions of users, exposure logs, and click logs. The test set includes millions of users, billions of exposure logs, and millions of click logs. The baselines encompass both DLRMs and GR models, including: DIN \cite{zhou2018deep}, which models short‑term user behavior with target attention; SIM(Soft) \cite{pi2020search}, a sequential model that uses soft‑search to encode user interests; TWIN \cite{si2024twin}, which extends multi‑head target attention from ESU to GSU; HSTU \cite{zhai2024actions}, an efficient model for long‑sequence behavior modeling; and LONGER \cite{chai2025longer}, a Transformer‑based architecture designed for ultra‑long behavior sequences. Experimental results are presented in Table~\ref{tab:overall_performance}: GRAB outperforms all other baselines,  achieving a 0.19\% relative improvement over the most competitive model. Meanwhile, Fig.~\ref{fig:overall} illustrates the performance of different models across varying lengths of user behavior sequences. GRAB surpasses other recommendation models at all sequence lengths, with its performance gains becoming more pronounced as the sequence length increases.

\begin{table}[htbp]
\centering
\small
\caption{Overall performance in industrial settings}
\label{tab:overall_performance}
\begin{tabular}{lc}
\toprule
Model & AUC  \\
\midrule
DIN & 0.83309 \\
SIM Soft & 0.83520 \\
TWIN & 0.83556 \\
HSTU & 0.83590 \\
\textbf{LONGER} & \textbf{0.83615} \\
\midrule
GRAB-small & 0.83661 \\
\textbf{GRAB-standard} & \textbf{0.83772} \\
\bottomrule
\end{tabular}
\end{table}

\subsection{Scaling Analysis}
We evaluate model performance across different capacity scales by independently scaling the number of Transformer blocks($n_{layer}$), the number of attention heads($n_{head}$), and the feature dimension of the model($d_{model}$) in Table~\ref{tab:scaling_config}, Fig.~\ref{fig:scaling} presents the test-set performance of the GRAB model under varying configurations (i.e., $n_{layer}$, $n_{head}$ and $d_{model}$). These results demonstrate that increasing model capacity effectively improves model performance. We also found that as the model capacity increases, the performance improvement on longer user behavior sequences becomes more significant. Moreover, no significant saturation trend is observed within the current range of configurations, which also confirms the strong scalability of the GRAB model.

\begin{table}[htbp]
\centering
\small 
\caption{Comparison of models with different settings}
\label{tab:scaling_config}
\begin{tabular}{lcc}
\toprule
Model & Params & Setting  \\
\midrule
GRAB$_{2l-2h-64d}$ & 6.51M & n$_{layer}$=2, n$_{head}$=2, d$_{model}$=64 \\
GRAB$_{4l-2h-64d}$ & 6.67M & n$_{layer}$=4, n$_{head}$=2, d$_{model}$=64 \\
GRAB$_{6l-2h-64d}$ & 6.83M & n$_{layer}$=6, n$_{head}$=2, d$_{model}$=64 \\
GRAB$_{2l-4h-64d}$ & 6.48M & n$_{layer}$=2, n$_{head}$=4, d$_{model}$=64 \\
GRAB$_{4l-4h-64d}$ & 6.63M & n$_{layer}$=4, n$_{head}$=4, d$_{model}$=64 \\
GRAB$_{4l-4h-128d}$ & 7.05M & n$_{layer}$=4, n$_{head}$=4, d$_{model}$=128 \\
GRAB$_{4l-4h-256d}$ & 8.13M & n$_{layer}$=4, n$_{head}$=4, d$_{model}$=256 \\
GRAB$_{4l-4h-512d}$ & 11.27M & n$_{layer}$=4, n$_{head}$=4, d$_{model}$=512 \\
\bottomrule
\end{tabular}
\end{table}

\subsection{Ablation Study}
\paragraph{Heterogeneous Tokens.} 
We conduct ablation studies on heterogeneous representations with three configurations: GRAB with heterogeneous, only partial, or only full tokens (Table~\ref{tab:hete_abla}). Results show that heterogeneous representations achieve the best performance. Using only partial tokens leads to significant degradation, confirming that full feature representations are more beneficial for target scoring. Notably, using only full tokens also degrades performance, suggesting that artificially designed statistical features can introduce confusion and impair sequence modeling.

\begin{table}[htbp]
    \centering
    \small 
    \caption{Ablation studies of GRAB}
    \label{tab:hete_abla}
    \begin{tabular}{lc}
        \toprule
        Model & AUC  \\
        \midrule
        \textbf{GRAB}  & 0.83772 \\
        \midrule
        GRAB w/ Partial Token &  0.83492 \\
        GRAB w/ Full Token &  0.83749 \\
        \midrule
        GRAB w/o relative pos &  0.83768\\
        GRAB w/o relative time &  0.83743 \\
        GRAB w/o relative action &  0.83724 \\
        \midrule
        GRAB w/o Multi-channel &  0.83743 \\
GRAB w/o Target-token mix &  0.83768\\
        \midrule
        \textbf{GRAB$_{sparse}$} & 0.83614 \\
        GRAB$_{sparse}$ w/o STS &  0.83549 \\
        \bottomrule
    \end{tabular}
\end{table}

\begin{figure}[htbp]
    \centering
    \includegraphics[width=0.40\textwidth]{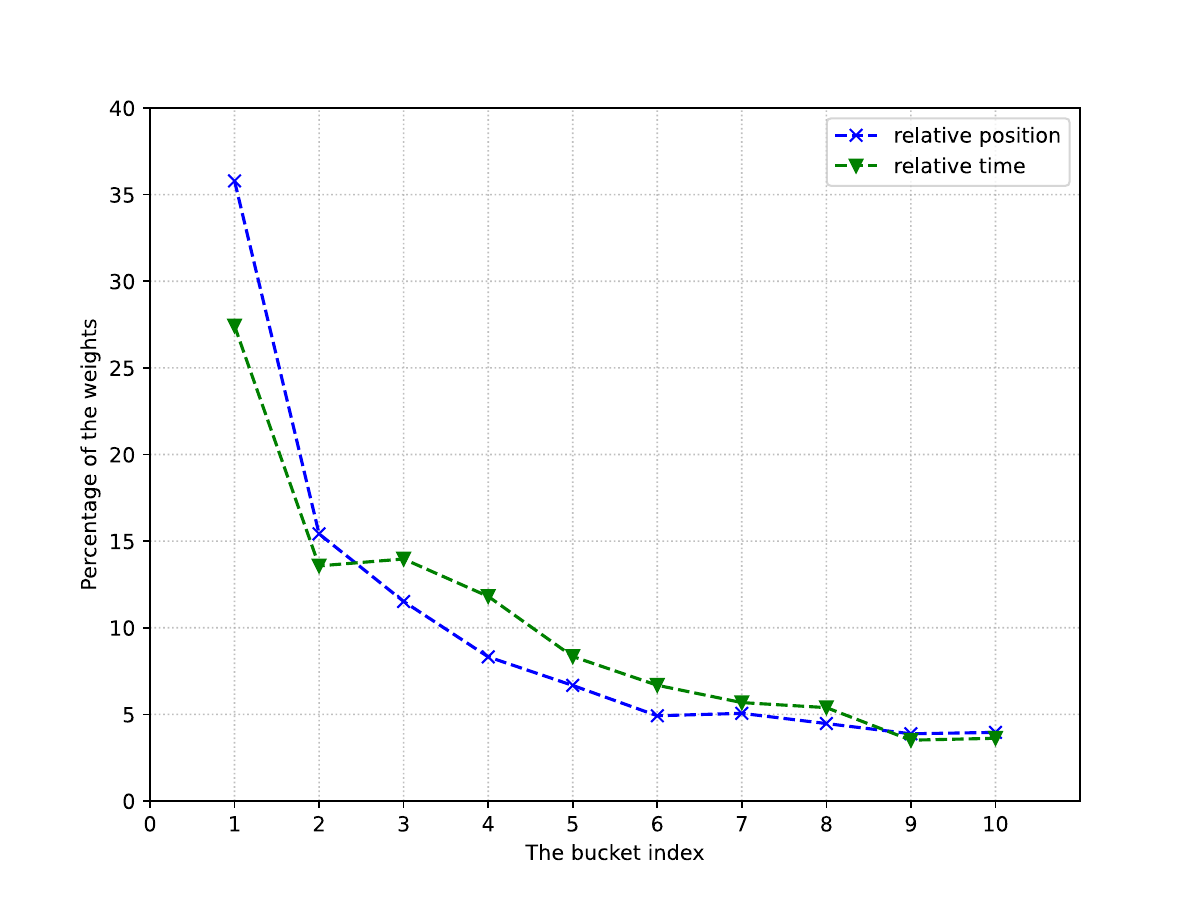} 
    \caption{The weight distribution of action-aware attention in relative postion and relative time.}
    \label{fig:action_weight_abla}
\end{figure}

\paragraph{Action-aware Attention.} 
We ablate three components of GRAB's Action-aware Attention: relative position, time, and action. The results (Table~\ref{tab:hete_abla}) show that removing any of these components degrades performance. The decline is more pronounced for time and action than for position, indicating that historical sequences are more sensitive to behavioral and temporal signals. We also analyze the attention weight distribution across buckets defined by relative position/time differences (smaller values denote more recent tokens). As shown in Figure~\ref{fig:action_weight_abla}, weights decrease as bucket values increase, confirming that more recent behaviors better reflect user interest and receive higher weights. For relative action, we compare positive (click) and negative (non‑click) labels. The weight distribution is highly skewed: positive labels account for 88\% of the total weight, versus only 12\% for negative labels. This suggests that incorporating more positive feedback could further improve sequence modeling.

\paragraph{Multi-channel Attention.} To verify the effectiveness of multi-channel attention in sequence modeling, we conduct the following settings: 1) the GRAB model without multi-channel attention, that is, using a single channel for sequence modeling, 2) remain the multi-channel attention and only remove the target token mix component. As shown in Table~\ref{tab:hete_abla}, both configurations have varying degrees of performance degradation, indicating that each component is indispensable. In terms of performance, multi-channel attention is crucial, and adding the target token mix component can further improve performance.

\paragraph{STS Training.}
We evaluate the STS paradigm by comparing GRAB's second-stage training with and without sequence modeling for sparse feature learning. With STS, sparse embeddings are updated through sequence modeling on packed user behavior sequences; without STS, the same batch data is treated as independent exposures. Results (as shown in Table~\ref{tab:hete_abla}) show that STS brings significant accuracy gains in sparse feature learning, confirming the efficacy of the two-stage training. This demonstrates that STS alleviates the distribution skew and overfitting caused by direct sequence-packed training.

\subsection{Online A/B Test}

To assess the online performance of GRAB, we deployed it in Baidu home feed scenario of Baidu and compared its performance with the current online DLRM model. The experiment used 10\% of the main traffic and remained online for about a month. Online evaluation shows that GRAB delivered 3.49\% improvement in CTR and 3.05\% improvement in CPM, which indicates that GRAB achieves more accurate advertising estimation and brings considerable revenue increments. Notably, GRAB has already been fully deployed on Baidu, and the online inference costs on par with the previous online DLRM model.

\section{Conclusion}

We propose GRAB, an end-to-end generative ranking framework that integrates a novel CamA mechanism to effectively capture temporal dynamics and specific action signals within user behavior sequences. On Baidu billion-scale industrial dataset, GRAB establishes a new state-of-the-art, outperforming DLRM and other GR baselines. Ablation studies validate the necessity of its key components, and our proposed STS training paradigm effectively mitigates distribution shift. Scaling analysis indicates continued gains from larger models and longer sequences. Finally, full online A/B testing in Baidu home feed ads shows that GRAB boosts CTR by 3.49\% and CPM by 3.05\%, leading to full production deployment. Further discussion of this work can be found in the Appendix~\ref{appendix:discussion}.

\nocite{langley00}

\bibliography{arxiv_grab}
\bibliographystyle{icml2026}

\newpage
\appendix
\onecolumn
\section{Extended Background}
\subsection{The Performance–Efficiency Trade-off in Industrial CTR Prediction}
\label{sec:performance_efficency}
The overall design of industrial-grade recommendation systems and their recommendation models almost always revolves around two goals: \textbf{performance} and \textbf{efficiency}\cite{covington2016deep,naumov2019deep,mudigere2022software,agarwal2023bagpipe,bai2025comprehensive}. Performance is not only reflected in the model's fitting capabilities as measured by metrics such as AUC and PCOC, but also in its ability to sensitively capture user interest drift and changes in content distribution under varying traffic patterns\cite{sheng2023joint,pi2020search}. Efficiency, on the other hand, is comprehensively reflected in the computational power consumption, memory/bandwidth usage, and online inference speed during the training and inference phases. Among these, training and inference costs are prerequisites for the long-term deployment and continuous iteration of the model in a real production environment\cite{naumov2019deep,mudigere2022software,agarwal2023bagpipe}. 

Although DLRMs have achieved considerable success, it faces bottlenecks in both performance and efficiency\cite{bai2025comprehensive,zhang2024wukong,han2025mtgr}. On the one hand, DLRMs rely on an experience- and rule-based feature system, which suffers from the inherent flaw of ``strong memory, weak reasoning.'' This makes the DLRMs insufficiently generalizable when dealing with new advertisements, new users, or scenarios requiring logical inference\cite{cheng2016wide,he2014practical,ma2018entire,ma2018modeling,wu2024survey}. At the same time, with the exponential growth of user behavior, traditional DLRMs suffer from significant information loss in ultra-long sequence modeling and has poor adaptability to different scenarios\cite{zhou2018deep,pi2020search,pi2019practice,zhang2024scaling}. On the other hand, as the network design of DLRMs become increasingly complex, the performance improvement of the model shows diminishing returns. To achieve further performance improvements, it often requires exponentially increased computational costs, making the long-term deployment and continuous iteration of the model in a real production environment problematic\cite{zhang2024wukong,zhang2024scaling,mudigere2022software,han2025mtgr}.

\subsection{A Taxonomy of LLM-based Recommendation Research}
\label{sec:LLM4rec}

LLMs have recently emerged as a promising direction for recommendation systems, giving rise to a growing line of research commonly referred to as LLM4Rec\cite{wu2024survey,lin2025can,zhao2024recommender,li2024large}. The motivation behind this paradigm shift lies in the inherent limitations of traditional ID-based recommendation models, which often struggle with semantic understanding, cold-start problems, and cross-domain generalization\cite{yuan2023go,li2025exploring}. LLMs offer the potential to introduce extensive world knowledge, robust reasoning capabilities, and high-quality textual generation into the recommendation pipeline\cite{zhang2025recommendation, he2023large}. However, integrating LLMs into industrial-scale systems presents unique challenges, primarily the ``ID-Text dilemma''---where high-cardinality sparse IDs do not map naturally to the continuous token space of LLMs\cite{tan2024idgenrec,rajput2023recommender}---and the prohibitive inference latency of decoder-only architectures in real-time scoring\cite{yue2023llamarec}. Based on recent literature and industrial practices, LLM4Rec approaches can be systematically categorized into three distinct paradigms: \emph{LLM as Recommender}, \emph{LLM for Representation}, and \emph{Generative Sequential Modeling}.

\textbf{LLM as Recommender.} This category explores the direct application of LLM capabilities---such as memory, reasoning, and zero-shot generalization---to core recommendation tasks including retrieval and ranking\cite{wu2024survey, lin2025can, xu2025tapping}. Methods in this domain typically adapt recommendation data into natural language prompts, leveraging techniques like Instruction Tuning to align the LLM with recommendation objectives\cite{zhu2024collaborative,zhang2025recommendation, bao2023tallrec,luo2025recranker}. While these methods demonstrate promise in explainability and conversational recommendation, their performance on traditional metrics (e.g., CTR) often falls short of specialized ID-based models\cite{liu2023chatgpt, di2023evaluating, cao2024aligning}. In recommendation scenarios, user behavior is heavily influenced by implicit feedback and specific context rather than the explicit semantic logic found in natural language; consequently, general world-knowledge reasoning does not necessarily translate effectively to modeling complex user--item interaction patterns\cite{bao2023tallrec, cao2024aligning, zhu2024collaborative}. Furthermore, the inference latency remains a significant bottleneck for real-time industrial deployment\cite{xu2025tapping}.

\textbf{LLM for Representation.} In this paradigm, LLMs function as sophisticated feature encoders \cite{lin2025can,wu2024survey}. Instead of performing the ranking directly, the intermediate layers or final output embeddings of the LLM are extracted and utilized as semantic features to augment the input of traditional recommendation models \cite{sun2024large, jia2025learn,geng2024breaking,chen2024hllm,ning2025user}. This approach aims to enhance the model's semantic understanding without bearing the full cost of LLM inference during the serving phase. LLM-derived representations significantly mitigate the limitations of discrete feature models, particularly regarding the generalization capability for long-tail items and cold-start users/ads \cite{hou2022towards,hou2023learning}. However, this methodology faces notable limitations. There is typically a limited gain on warm items, as the strong collaborative filtering signals derived from abundant historical interactions often outweigh the semantic benefits provided by the LLM \cite{hou2023learning,lin2024pre}. Furthermore, employing large-scale models for representation learning introduces a high inference cost, which creates substantial latency and resource bottlenecks during both the offline feature extraction and online serving phases \cite{lin2025can}.

\textbf{Generative Sequential Modeling.} This category represents a structural adaptation rather than a direct semantic application. It borrows the architectural innovations underlying LLMs---specifically the Transformer architecture, Causal Masking, and Long-context modeling capabilities---to reconstruct recommendation systems \cite{vaswani2017attention,kang2018self,sun2019bert4rec}. These models(such as GR models) treat user history as a sequence and the next item prediction as a generative task, similar to next token prediction \cite{kang2018self,petrov2023generative,han2025mtgr}. By employing generative sequential modeling techniques and combining them with discrete features that precisely characterize user historical behavior, these models have shown significant potential \cite{han2025mtgr}. A key observation in this domain is the emergence of ``scaling laws'' within recommendation systems, where model performance metrics improve predictably as the sequence length and model capacity increase \cite{shin2023scaling,zhang2024scaling}, mirroring the trajectory seen in NLP.

A comparative overview of the three LLM4Rec paradigms is presented in Table~\ref{tab:llm4rec_comparison}, highlighting their core mechanisms, key strengths, and primary limitations.

\begin{table}[h]
\centering
\caption{Comparison of LLM4Rec Paradigms.}
\label{tab:llm4rec_comparison}
\resizebox{\textwidth}{!}{%
\begin{tabular}{l l l l}
\toprule
\textbf{Category} & \textbf{Core Mechanism} & \textbf{Key Strength} & \textbf{Primary Limitation} \\
\midrule
\textbf{LLM as Recommender} & \begin{tabular}[c]{@{}l@{}}Instruction tuning / \\ prompting for ranking\end{tabular} & \begin{tabular}[c]{@{}l@{}}Explainability, \\ Zero-shot generalization\end{tabular} & \begin{tabular}[c]{@{}l@{}}High latency, \\ ineffective interaction modeling\end{tabular} \\
\midrule
\textbf{LLM for Representation} & \begin{tabular}[c]{@{}l@{}}Feature Encoder \\ (Extracting embeddings)\end{tabular} & \begin{tabular}[c]{@{}l@{}}Cold-start handling, \\ Semantic understanding\end{tabular} & \begin{tabular}[c]{@{}l@{}}Limited gain on warm items,\\ high inference cost\end{tabular} \\
\midrule
\textbf{Generative Seq. Modeling} & \begin{tabular}[c]{@{}l@{}}Transformer decoder, \\ next token prediction\end{tabular} & \begin{tabular}[c]{@{}l@{}}Scaling laws,\\ high capacity\end{tabular} & \begin{tabular}[c]{@{}l@{}}High inference latency, \\ high resource consumption\end{tabular} \\
\bottomrule
\end{tabular}
}
\end{table}

\subsection{Limitations of GR in Performance}
\label{appendix:lack_of_performance}
Existing GR models often inherit NLP-style causal Transformers with minimal adaptation to recommender logs, implicitly assuming that user history can be represented as a homogeneous token stream. In practice, recommendation data are inherently \emph{heterogeneous}: events comprise multiple fields (e.g., item, context, query, creator), and user trajectories interleave distinct behavior types (e.g., exposure, click, like, skip, dwell). A naive serialization pipeline typically collapses this structured record into a single sequence of item IDs (or flattened tokens), which \emph{discards action semantics}—the critical distinction between what the user was shown and how the user responded.

This structural mismatch leads to a performance bottleneck: the model conflates semantically different interactions, dilutes supervision signals, and learns spurious correlations (e.g., treating exposures as implicit positives or mixing weak/strong feedback). As a result, even with larger models and longer contexts, GR may underperform in industrial CTR settings where fine-grained behavior semantics and cross-field interactions are decisive, highlighting the need for action-aware, heterogeneity-preserving sequence modeling rather than direct NLP-style tokenization.

\section{Extended Related Work}
\label{sec:appendix_gr}
\subsection{Limitations of Emerging Generative Ranking Models}

While GR models have successfully introduced the scaling laws of LLMs into recommendation systems, their direct application to industrial CTR prediction faces distinct structural and optimization challenges.

\textbf{Mitigation of Distribution Skew in Sequence Packing.} To improve training efficiency with variable-length user sequences, standard GR models often employ sequence packing techniques borrowed from NLP (e.g., concatenating multiple short sequences). However, unlike NLP where samples are generally Independent and Identically Distributed (I.I.D.), packing in recommendation systems groups multiple interactions from the \textit{same user} into a single training instance to maintain context. This creates a severe distribution skew, where a mini-batch is dominated by highly correlated samples from a few users. This correlation causes the model—especially the sparse embedding parameters—to overfit specific user identities rather than learning generalizable interaction patterns.

\textbf{Action heterogeneity.} Existing GR models often treat user history as a homogeneous token stream, neglecting the inherent heterogeneity of recommendation data. This reliance on naive serialization discards critical action semantics—distinguishing what was shown from how the user responded—thereby diluting supervision signals and limiting performance in complex industrial scenarios (as discussed in Appendix~\ref{appendix:lack_of_performance}).

\textbf{Explicit Modeling of Relative Action Signals.} Standard GR models rely on the vanilla Scaled Dot-Product Attention, often supplemented only by absolute or relative positional encodings. While effective for capturing sequential order (as emphasized in LONGER), this approach treats the {nature} of the interaction as implicit. It fails to explicitly differentiate between varying feedback signals (e.g., ``clicked'' vs. ``viewed'' vs. ``purchased'') and their relative timing in a query-dependent manner.

\subsection{Comparative Discussion with Existing Ranking Models}
To position GRAB within the evolving landscape of recommendation systems, we provide a qualitative comparison against two primary categories of existing models: traditional DLRMs and emerging GR approaches in Table~\ref{tab:comparison}.

\begin{table*}[htbp]
\centering
\caption{Comparison of DLRM, HSTU, LONGER, and GRAB based on Key Dimensions from the Paper}
\label{tab:comparison}
\resizebox{\textwidth}{!}{
\begin{tabular}{c c c c c}
\toprule
\textbf{Dimension} & \textbf{DLRMs} & \textbf{HSTU} & \textbf{LONGER} & \textbf{GRAB} \\ \midrule
\textbf{Core Architecture} & \begin{tabular}[c]{@{}c@{}}Embedding + MLP + \\ Target Attention (DIN/SIM)\end{tabular} & \begin{tabular}[c]{@{}c@{}}Pure Transformer \\ (Generative Ranking)\end{tabular} & \begin{tabular}[c]{@{}c@{}}Transformer-based \\ (Optimized for Long Seq)\end{tabular} & \begin{tabular}[c]{@{}c@{}}\textbf{End-to-End Generative Framework} \end{tabular} \\ \midrule
\textbf{Feature Interaction} & \begin{tabular}[c]{@{}c@{}}Manual Cross Features \\ \& MLP / Gating\end{tabular} & \begin{tabular}[c]{@{}c@{}}Standard Self-Attention \\ (Global)\end{tabular} & \begin{tabular}[c]{@{}c@{}}Standard Self-Attention \\ (Long Context)\end{tabular} & \begin{tabular}[c]{@{}c@{}}\textbf{Causal Action-aware} \\ \textbf{Multi-channel Attention (CamA)}\end{tabular} \\ \midrule
\textbf{Heterogeneous Features} & \begin{tabular}[c]{@{}c@{}}Concatenation \& Flattening \\ (Fixed-length vector)\end{tabular} & \begin{tabular}[c]{@{}c@{}}Homogeneous Tokens \\ (Often redundant static info)\end{tabular} & \begin{tabular}[c]{@{}c@{}}Homogeneous Tokens \\ (Similar to HSTU)\end{tabular} & \begin{tabular}[c]{@{}c@{}}\textbf{Heterogeneous Tokens} \\ \textbf{(Partial History / Full Candidate)}\end{tabular} \\ \midrule
\textbf{Action Semantic Modeling} & \begin{tabular}[c]{@{}c@{}}Implicit or via \\ Engineered Features\end{tabular} & \begin{tabular}[c]{@{}c@{}}Implicit \\ (Learned via sequence order)\end{tabular} & \begin{tabular}[c]{@{}c@{}}Implicit \\ (Focus on sequence length)\end{tabular} & \begin{tabular}[c]{@{}c@{}}\textbf{Action-aware RAB} \\ \textbf{(Explicit Relative Bias)}\end{tabular} \\ \midrule
\textbf{Training Strategy} & \begin{tabular}[c]{@{}c@{}}Standard Supervised \\ (Pointwise BCE)\end{tabular} & \begin{tabular}[c]{@{}c@{}}Autoregressive / Next-Token \\ (Suffers from Dist. Skew)\end{tabular} & \begin{tabular}[c]{@{}c@{}}Autoregressive \\ (Standard GR Training)\end{tabular} & \begin{tabular}[c]{@{}c@{}}\textbf{Sequence-Then-Sparse (STS)} \\ \textbf{(Decoupled Optimization)}\end{tabular} \\ \midrule
\textbf{Scaling Potential} & \begin{tabular}[c]{@{}c@{}}Diminishing Returns \\ (Saturated Performance)\end{tabular} & \begin{tabular}[c]{@{}c@{}}High \\ (Follows Scaling Laws)\end{tabular} & \begin{tabular}[c]{@{}c@{}}High \\ (Scaling in Sequence Length)\end{tabular} & \begin{tabular}[c]{@{}c@{}}\textbf{High \& Linear} \\ \textbf{(No saturation observed)}\end{tabular} \\ \bottomrule
\end{tabular}}
\end{table*}

\section{Detailed Architecture and Data Flow of GRAB}
\label{detail_GRAB}
\subsection{Sparse Feature Layer}
\label{sec:sparse}
In the sparse feature layer, GRAB expands user $u$'s raw behavior sequence and the candidate ad sequence into event-level representations, preserving their original temporal order to form a structured, time-ordered event sequence:
\begin{equation}
\label{eq:seq}
\left\{ S_t^{beh} \right\}_{t=1}^{T_u},\qquad
\left\{ S_i^{ad} \right\}_{i=1}^{N_u}
\end{equation}
where $T_u$ denotes the length of user $u$'s behavior history, $N_u$ denotes the number of candidate advertisements for user $u$. GRAB takes user behavior history and candidate ads as input sequences. Specifically, the $t$-th behavior event $S_t^{beh}$ consists of user attributes $U$, context $C$, and behavior attributes $B$, while the $i$-th candidate ad $S_i^{ad}$ consists of context $C$ and item attributes $A$, as follows:
\begin{equation}
S_t^{beh}=\big(U_u, C_t, B_t\big), \quad S_i^{ad}=\big(A_i, C_i\big)
\end{equation}

Following the discrete feature engineering standard of DLRMs, we apply a structured expansion function $\Phi$ to transform each event into a fixed multi-field representation. Subsequently, each field value is mapped to a discrete ID via a sparse PSTable $\Pi$. The event-level representations of the raw behavior sequence and the candidate ad sequence can be obtained as:
\begin{equation}
\begin{aligned}
\mathbf{x}_t^{beh} &= \Pi\big(\Phi(S_t^{beh})\big) , \quad &t=1,\ldots,T_u \\
\mathbf{x}_i^{ad} &= \Pi\big(\Phi(S_i^{ad})\big) , \quad &i=1,\ldots,N_u
\end{aligned}
\end{equation}

\subsection{Dense Tokenizer}
\label{sec:dense}
Unlike the DLRM approach, which concatenates all field embeddings into a fixed-length flattened vector, GRAB preserves the event structure by aggregating field embeddings within each event into a single event token. This yields a time-ordered token sequence that is fed into a Transformer to capture long-range behavioral dependencies and interest drift. Given the structured discrete ID sequences from Section~\ref{sec:sparse}, GRAB converts each event into a dense token for Transformer-based sequential modeling. 
Specifically, each event is first transformed into a vector through a field-wise embedding lookup followed by a multi-field fusion process, as follows:
\begin{equation}
\begin{gathered}
v_{t}=\mathrm{Emb}\!\left(x_{t}\right)
,\quad
v_{i}=\mathrm{Emb}\!\left(x_{i}\right)
\end{gathered}
\end{equation}
Subsequently, the token representation for each event is generated by the GateMLP module, which consists of an MLP and a Gate layer, formulated as:
\begin{equation}
\begin{gathered}
\mathbf{e}=\text{GateMLP}\big(v)\in\mathbb{R}^{d_{model}}\\
\mathbf{E}^{beh}=\{\mathbf{e}^{beh}_t\}_{t=1}^{T_u}
, \qquad
\mathbf{E}^{ad}=\{\mathbf{e}^{ad}_i\}_{i=1}^{N_u}.
\end{gathered}
\end{equation}

\subsection{Autoregressive-like Sequence Modeling Layer}
\label{sec:appendix_seq}
Taking the output ($\mathbf{E}^{beh}$ and ${E}^{ad}$) from the previous dense layer as input, we feed it into the sequence modeling layer(for more details about this layer, please see Section~\ref{sec:layer_sequence}) to capture sequential dependencies. Let $Z$ denote the final output representation of the sequence layer:
\begin{equation}
Z = \operatorname{SeqLayer}(\mathbf{E}^{beh}, \mathbf{E}^{ad}).
\label{eq:seq_layer}
\end{equation}

Finally, the output of the sequence modeling layer is fed into a logistic head to yield the CTR prediction. The model is trained to minimize the binary cross-entropy loss:
\begin{equation}
\label{eq:seq_CTR_head}
\begin{gathered}
\hat{y}=\sigma\big(\mathbf{w}^\top Z+b\big)\in(0, 1),\\
\mathcal{L}_{\mathrm{BCE}}=-\big[y\log\hat{y}+(1-y)\log(1-\hat{y})\big].
\end{gathered}
\end{equation}

\section{In-Depth Analysis of STS Training}
\label{appendix:STS}
While sequence packing dramatically improves computational resource utilization by eliminating padding, it introduces a non-trivial optimization challenge known as Distribution Skew. In this section, we provide the theoretical justification for the proposed STS training paradigm, detail its mathematical formulation, and discuss how it reconciles the learning space inconsistency between stages.

\subsection{Distribution Skew}
\label{appendix:skew}
Standard Stochastic Gradient Descent (SGD) relies on the assumption that samples within a mini-batch are I.I.D.. Formally, for a loss function $\mathcal{L}$, the gradient $g$ computed on a batch $\mathcal{B}$ is an unbiased estimator of the true gradient:
\begin{equation}
\mathbb{E}[g_{\mathcal{B}}] = \nabla \mathcal{L}, \quad \text{Var}(g_{\mathcal{B}}) = \frac{\sigma^2}{|\mathcal{B}|}
\end{equation}
where $|\mathcal{B}|$ is the batch size.

In sequence packing, we form a packed mini-batch $\mathcal{B}_{\text{pack}}$ by concatenating multiple actions from the same user $u$ into a single training instance (or a user-dominated batch) to avoid padding waste. While efficient, this construction makes samples within $\mathcal{B}_{\text{pack}}$ highly correlated: for $i,j \in \mathcal{B}_{\text{pack}}$, $\mathrm{Cov}(x_i,x_j)\gg 0$ , because they share the same user\_id, static context, and long-term interests. As a result, the effective batch size is substantially reduced and the variance of the stochastic gradient estimator increases, yielding noisier and less stable updates.

This issue is most damaging for the sparse embedding table $\Phi$. Since a packed batch repeatedly contains the same user features (e.g., {user\_id=123} appears $L$ times along the packed sequence), the update for that single embedding vector is amplified by repeated contributions:

\begin{equation}
\Delta \Phi_{u} \propto \sum_{t=1}^{L} \nabla \mathcal{L}_t .
\end{equation}

Such oversized, user-specific updates encourage $\Phi$ to {memorize} individual trajectories rather than learn generalizable interaction patterns. Meanwhile, the dense sequence model (e.g., Transformer) suffers from batch-to-batch distribution skew: consecutive packed batches may be dominated by different users (User A $\rightarrow$ User B), causing abrupt shifts in inputs and gradients, which hinders stable convergence of sequential reasoning parameters.

\subsection{Formalization of STS Stages}
To mitigate the distribution  skew, STS decouples the optimization into two orthogonal objectives: relational reasoning (Dense) and feature representation (Sparse). The algorithm flow of STS is shown in Algorithm~\ref{alg:training_stages}.

\begin{algorithm}[htbp]
\caption{Two-Stage Alternating Training Strategy}
\label{alg:training_stages}
\begin{algorithmic}[1]
\REQUIRE Training dataset $\mathcal{D}$, Learning rate $\eta$
\STATE \textbf{Initialize:} Dense tokenizer $\theta_{cont}$, Causal Transformer $\theta_{tr}$, Sparse Embedding Table $\Phi$

\WHILE{not converged}
    \STATE \textbf{// Stage I: Sequence Modeling (Sequence Phase)}
    \STATE Sample packed user sequences batch $Z = (x_{hist}, x_{cand}) \sim \mathcal{D}$
    \STATE \textit{Freeze} Sparse Embedding Table $\Phi$
    \STATE Compute sequence output: 
    \STATE \quad $\hat{y}_{seq} \leftarrow f_{seq}(Z; \theta_{cont}, \theta_{tr}, \Phi)$
    \STATE Compute sequence loss:
    \STATE \quad $\mathcal{L}_{seq} \leftarrow \text{BCE}(\hat{y}_{seq}, y)$
    \STATE Update dense modules:
    \STATE \quad $\{\theta_{cont}, \theta_{tr}\} \leftarrow \text{SGD}(\nabla_{\{\theta_{cont}, \theta_{tr}\}} \mathcal{L}_{seq})$
    
    \STATE \textbf{// Stage II: Sparse Feature Learning (Sparse Phase)}
    \STATE Sample independent user-ad batch $(x^{beh}, x^{ad}) \sim \mathcal{D}$
    \STATE \textit{Freeze} Sequential modules $\{\theta_{cont}, \theta_{tr}\}$
    \STATE Compute aggregated features (breaking distribution skewness):
    \STATE \quad $s \leftarrow \text{Agg}(\{\Phi(x_t^{beh})\}) \parallel \Phi(x^{ad})$
    \STATE Compute sparse phase prediction:
    \STATE \quad $\hat{y}_{sp} \leftarrow f_{sp}(s)$
    \STATE Update sparse embeddings:
    \STATE \quad $\Phi \leftarrow \text{SGD}(\nabla_{\Phi} \mathcal{L}_{sp})$
\ENDWHILE
\end{algorithmic}
\end{algorithm}

\subsection{Discussion: A Pre-training \& Transfer Perspective}
The STS paradigm can be viewed through the lens of pre-training and transfer learning. Stage I serves as a sequential pre-training step that encodes interest evolution into the dense space, while Stage II transfers these insights back to the target sparse feature space for fine-tuning.
Although this non-end-to-end mode might introduce a subtle objective inconsistency between stages, our results demonstrate that the benefit of resolving distribution skew far outweighs the cost of misalignment. STS ensures that the end-to-end sequence predictor (Stage I) consistently validates against the optimal embeddings refined in Stage II, providing a pragmatic path to balance efficiency and generalization in large-scale recommendation systems.

\section{System Deployment Details}
\label{sec:appendix_deployment}

\subsection{Platform Architecture}
\begin{figure*}[t]
    \centering
    \includegraphics[width=0.8\linewidth]{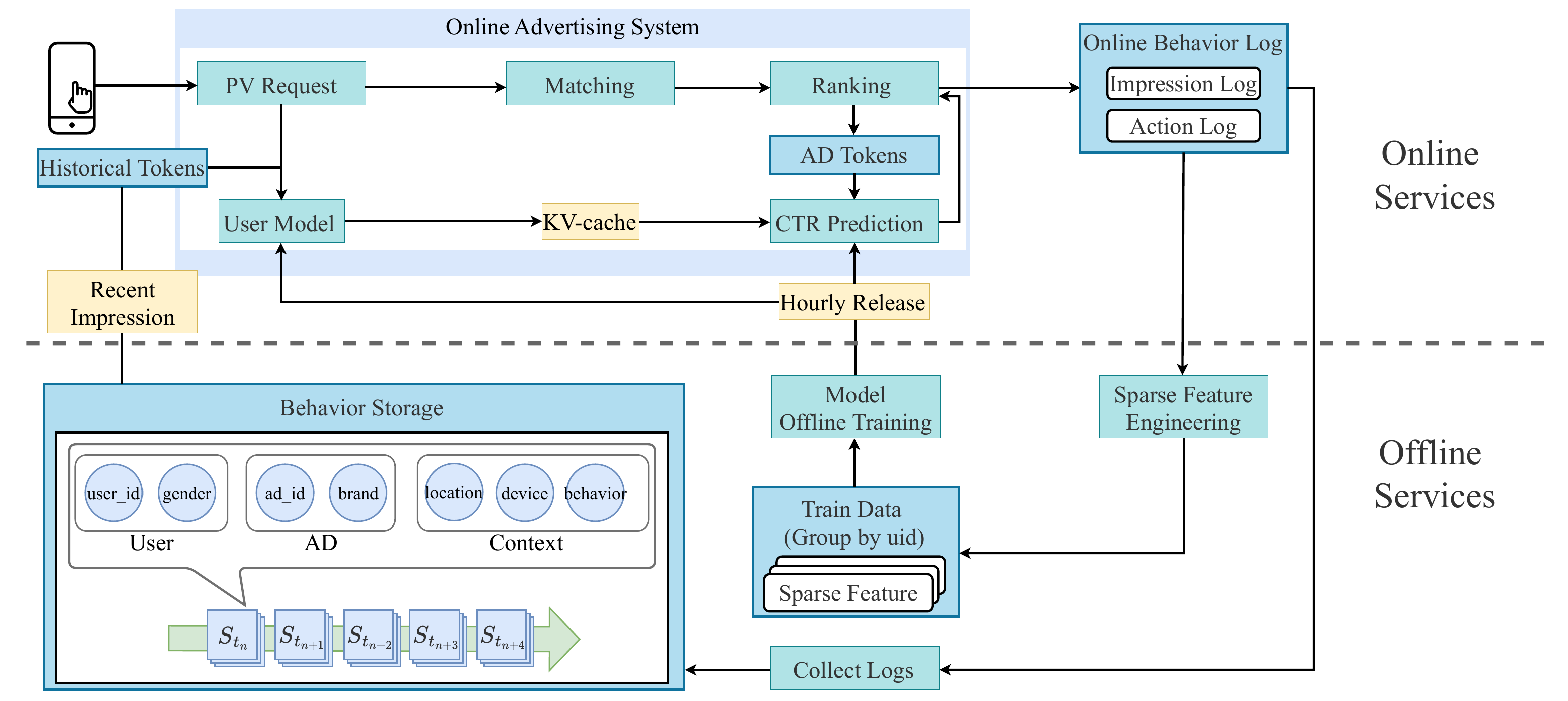}
    \caption{Overview of an online advertising CTR system with an online--offline closed loop. Online services handle PV requests via matching and ranking, and feed the CTR predictor with user-side historical tokens (maintained by a user model with KV-cache) and candidate ad tokens; user interactions are continuously logged as impression/action logs. Offline services collect these logs, apply sparse feature engineering, group training samples by user ID, and perform offline training; updated models are released (e.g., hourly) back to online.}
    \label{fig:pipeline}
\end{figure*}

As illustrated in Fig.~\ref{fig:pipeline}, the proposed system is implemented within a comprehensive Online Advertising System that operates as a closed-loop platform integration of Online Services and Offline Services. The architecture is designed to handle high-concurrency requests while maintaining model freshness through continuous updates.

\subsubsection{Online Serving}
The online serving component processes user interactions in real-time. The workflow initiates with a Page View (PV) Request, which sequentially passes through \emph{matching} and \emph{ranking} phases to select appropriate advertisements. 

The core of the ranking mechanism involves a CTR Prediction module that relies on two primary inputs:
\begin{itemize}
    \item \textbf{User Representation:} A user model processes historical tokens and maintains a KV-cache to efficiently manage state.
    \item \textbf{Ad Representation:} The system generates ad tokens corresponding to candidate advertisements.
\end{itemize}
This process results in the display of ads, where user interactions are captured in the online behavior log, consisting of impression logs and action logs.

\subsubsection{Offline Training and Feedback Loop}
The offline component ensures the model evolves with user behavior. The process involves:
\begin{itemize}
    \item \textbf{Data Collection and Storage:} Logs are collected and stored in behavior storage, which organizes data into User (user\_id, gender), AD (ad\_id, brand), and Context (location, device, behavior) categories.
    \item \textbf{Feature Engineering:} The system performs sparse feature engineering on the collected logs.
    \item \textbf{Training:} Training data is grouped by user ID (uid) to facilitate model offline training.
    \item \textbf{Deployment:} Updated models are pushed back to the online environment via an hourly release mechanism, updating the CTR prediction and user model components.
\end{itemize}

\subsection{Deployment Constraints and Optimization}
The system, referred to as GRAB, has been deployed in a feed ad ranking system handling billions of daily requests. The deployment addresses several critical engineering challenges distinguishable from conventional DLRMs:

\begin{itemize}
    \item \textbf{Computational Complexity:} Unlike memory-bound DLRMs, GR in this context is compute-bound. This is primarily attributed to the quadratic complexity ($O(L^2)$) of Transformer self-attention mechanisms required for processing long sequences.
    \item \textbf{Latency Requirements:} The system's performance is bound by critical latency thresholds defined in the Service Level Agreements (SLAs).
\end{itemize}

To meet these demands, the deployment utilizes a co-designed architecture incorporating data compression, hierarchical storage, and disaggregated serving.

\subsection{Data Infrastructure and Training Optimization}

\textbf{User-Centric Data Layout and Compression.} Traditional industrial pipelines use time-partitioned logs, necessitating expensive global shuffling. We transitioned to a User Slice storage architecture, where user behavior logs are pre-aggregated by User ID into contiguous physical file blocks. To further reduce I/O overhead, we upgraded from standard Gzip text storage to a Binary + LZ4 compression scheme. The binary format combined with LZ4 (which is highly efficient for repetitive user behavior patterns) reduced storage costs by 12\% compared to Gzip and decreased decoding latency by 70\%, enabling the system to stream complete user histories with near-linear scalability.

\textbf{Hierarchical Parameter Server (PaddleBox).} To handle terabyte-scale embedding tables, we utilized the training framework with a three-tier storage hierarchy:
\begin{itemize}
    \item \textbf{L1 (GPU HBM):} Stores hot embeddings for the current micro-batch.
    \item \textbf{L2 (CPU DRAM):} Buffers warm parameters.
    \item \textbf{L3 (SSD):} Utilizes NVMe SSDs for massive long-tail feature embeddings. An intelligent prefetching engine asynchronously moves parameters between tiers, masking SSD I/O latency.
\end{itemize}

\textbf{Handling Long-Tail Sequences.} Real-world user history lengths exhibit a heavy-tailed distribution, where the top 5\% of sequences can cause ``Out Of Memory'' (OOM). We implemented an Inverse Sliding Window strategy during training. Instead of random slicing, sequences are sliced from the most recent action backwards. This prioritizes recent user interests and ensures that extreme long-tail data does not destabilize GPU memory usage.

\subsection{High-Performance Inference Architecture}
\label{appendix:high_performance}
\textbf{Disaggregated Serving and Parallelism.} We adopted a disaggregated serving architecture using a User Interest Center (UIC). The UIC asynchronously computes and updates the Transformer's Key-Value (KV) cache triggered by user actions. Crucially, we implemented Parallel Material Recall, where the user's historical sequence encoding overlaps with the candidate generation (ad retrieval) phase. By the time candidate ads are retrieved, the user's dense state is already computed, significantly hiding latency.

\textbf{KV-Cache Reuse and M-FALCON.} To avoid re-computing the user history for every candidate ad, we integrated the M-FALCON algorithm \cite{zhai2024actions}. It utilizes a broadcast-attention mechanism where the fetched user KV cache is shared across a micro-batch of candidate items, reducing the marginal inference complexity per item from quadratic to linear.

\textbf{Operator Fusion and Mixed Precision.} To maximize throughput on GPUs, we employed aggressive operator fusion (e.g., fusing Gemm + Bias + LayerNorm), which reduced kernel launch overheads and improved latency by roughly 43\%. Furthermore, rather than simple INT8 quantization which may degrade accuracy, we adopted a Mixed Precision strategy: utilizing TF32 for Transformer matrix operations to accelerate computation and FP16 for fully connected layers, achieving a balance between 28\% performance gain and negligible precision loss.

\subsection{Data Consistency}
A major challenge in GRs is the Freshness Gap (Train-Serve Skew). We addressed this by implementing a streaming data pipeline based on Flink \& TableStore. We utilized a Global Strictly Incremental ID mechanism to ensure strict ordering of user actions across distributed nodes. This allows the inference engine to fetch the exact state of the user corresponding to the training checkpoint, reducing data synchronization delay from minutes to seconds and ensuring the model always predicts based on the most consistent context.

\section{Discussion}
\label{appendix:discussion}
\subsection{Limitations and Challenges}
\textbf{Operational Complexity of Two-Stage Training.} A primary limitation of GRAB lies in the operational overhead introduced by the Sequence Then Sparse (STS) training paradigm. While STS effectively resolves the distribution skew caused by sequence packing and stabilizes the optimization of dense versus sparse parameters, it inherently complicates the model iteration pipeline. Unlike standard DLRMs that support continuous, single-stage online learning, STS requires a decoupled scheduling of sequence modeling (freezing sparse features) and sparse feature learning (freezing dense parameters). This increases the engineering maintenance cost and introduces latency in incorporating fresh feature embeddings into the dense sequential context, potentially affecting the model's responsiveness to emerging trends in real-time environments.

\textbf{Compute-Bound Hardware Constraints.} The shift from DLRMs to Generative Ranking marks a fundamental transition from memory-bound to compute-bound workloads in recommendation infrastructure. Although we mitigated inference latency via optimizations like Action-aware RAB and KV-cache reuse (Appendix~\ref{appendix:high_performance}), the quadratic complexity of attention mechanisms—even when bounded by sliding windows—remains computationally heavier than the MLP layers of traditional models. Scaling GRAB to significantly longer sequences (e.g., user lifetimes spanning months) or deploying it on edge devices with limited compute capacity poses a significant challenge, necessitating further research into linear attention mechanisms or more aggressive token pruning strategies specifically tailored for recommendation data.

\textbf{Interpretability of Generative Signals.} While the Multi-channel Attention mechanism allows us to inspect which behavior subsets contribute to a prediction, the end-to-end generative nature of GRAB can obscure the precise ``why'' behind a ranking decision compared to feature-engineered linear models. Understanding whether a prediction is driven by short-term intent (sequential reasoning) or long-term habit (sparse memorization) remains a non-trivial task, which is critical for debugging bad cases in commercial systems.

\subsection{Future Directions}
\textbf{Towards Multimodal Generative Ranking.} Currently, GRAB operates on discretized ID tokens derived from categorical features. However, the architecture's Heterogeneous Token design (Section~\ref{sec:het}) is naturally extensible to other modalities. A promising future direction is to integrate raw modal tokens—such as image patches (Visual Tokens) or ad textual descriptions (Language Tokens)—directly into the interaction sequence. By leveraging GRAB’s strong sequence modeling capabilities, the model could learn semantic alignment between visual cues and user behaviors end-to-end, overcoming the information loss inherent in pre-extracted ID features and enabling true multimodal CTR prediction.

\textbf{Unified Generative Representation Across Domains.} Finally, the ``pre-training \& fine-tuning'' paradigm common in NLP has yet to be fully realized in industrial recommendation. We envision extending GRAB to learn a Universal User Representation by pre-training on diverse behavior logs across multiple business scenarios (e.g., Home Feed, Search, and Short Video). A unified GRAB model could transfer learned sequential patterns from data-rich domains to cold-start scenarios, effectively solving the ``data silo'' problem prevalent in large-scale platforms.

\textbf{Foundation for Agent-based Recommender Systems.} GRAB’s ability to model the transition probabilities of user states ($s_t \xrightarrow{} s_{t+1}$) positions it as a powerful ``World Model'' or User Simulator for future agent-based recommendation systems. By accurately predicting not just the next click, but the evolution of user interests over time, GRAB can serve as the environment model for Reinforcement Learning (RL) agents. This would allow the system to move beyond myopic CTR optimization toward maximizing Long-Term Value (LTV) or user satisfaction by simulating how current recommendations influence future user trajectories.

\end{document}